\documentclass[english,aps,manuscript,showpacs,preprint,superscriptaddress]{revtex4-1}%
\usepackage{amsmath}
\usepackage{amssymb}
\usepackage{graphicx}

\makeatletter

\providecommand{\tabularnewline}{\\}

\usepackage{amsfonts}\usepackage{amsthm}\usepackage{latexsym}
\usepackage{color}
\usepackage{xcolor}\usepackage{subfigure}\usepackage{manfnt}
\usepackage{picinpar}
\usepackage{subeqnarray}
\usepackage{cases}
\usepackage{extarrows}
\usepackage{epstopdf}
\usepackage{picins}

\newcommand{\rom}[1]{\uppercase\expandafter{\romannumeral#1}}

\newcommand{\beq}{\begin{equation}}
\newcommand{\eeq}{\end{equation}}
\newcommand{\bal}{\begin{align}}
\newcommand{\eal}{\end{align}}
\newcommand{\baln}{\begin{align*}}
\newcommand{\ealn}{\end{align*}}

\graphicspath{{fig/}}

\begin{document}

\title{High order  volume-preserving algorithms for relativistic charged particles in general electromagnetic fields}

\author{Yang He }

\affiliation{Department of Modern Physics and Collaborative Innovation Center
for Advanced Fusion Energy and Plasma Sciences, University of Science
and Technology of China, Hefei, Anhui 230026, CHINA}

\affiliation{Key Laboratory of Geospace Environment, CAS, Hefei, Anhui 230026,
CHINA}

\author{Yajuan Sun}

\affiliation{LSEC, Academy of Mathematics and Systems Science, Chinese Academy
of Sciences, P.~O.~Box 2719, Beijing 100190, CHINA}

\author{Ruili Zhang}

\affiliation{Department of Modern Physics and Collaborative Innovation Center
for Advanced Fusion Energy and Plasma Sciences, University of Science
and Technology of China, Hefei, Anhui 230026, CHINA}

\affiliation{Key Laboratory of Geospace Environment, CAS, Hefei, Anhui 230026,
CHINA}

\author{Yulei Wang}

\affiliation{Department of Modern Physics and Collaborative Innovation Center
for Advanced Fusion Energy and Plasma Sciences, University of Science
and Technology of China, Hefei, Anhui 230026, CHINA}

\affiliation{Key Laboratory of Geospace Environment, CAS, Hefei, Anhui 230026,
CHINA}

\author{Jian Liu}

\affiliation{Department of Modern Physics and Collaborative Innovation Center
for Advanced Fusion Energy and Plasma Sciences, University of Science
and Technology of China, Hefei, Anhui 230026, CHINA}

\affiliation{Key Laboratory of Geospace Environment, CAS, Hefei, Anhui 230026,
CHINA}

\author{Hong Qin}
\affiliation{Department of Modern Physics and Collaborative Innovation Center
for Advanced Fusion Energy and Plasma Sciences, University of Science
and Technology of China, Hefei, Anhui 230026, CHINA}
\affiliation{Plasma Physics Laboratory, Princeton University, Princeton, New Jersey
08543, USA}



\begin{abstract}
We construct high order symmetric volume-preserving methods for the relativistic
dynamics of a charged particle by the splitting technique with processing.
Via expanding the phase space to include time $t$, we give a more general construction
 of volume-preserving methods that can be applied to systems with time-dependent
electromagnetic fields. The newly derived methods provide  numerical solutions with good accuracy and conservative properties over long time of simulation. Furthermore, because of the use of processing technique the high order methods
are explicit, and cost less than the methods derived from standard compositions,
thus are more efficient. The results are verified by the numerical
experiments. Linear stability analysis of the methods show that the
high order processed method allows larger time step size during integration.
\end{abstract}


\maketitle

\section{Introduction}

The dynamics of relativistic particles under the influence
of electromagnetic fields is a fundamental process in  plasma physics, space physics, accelerator physics, etc..
Numerical simulations on trajectories of charged particles have been widely used to study their dynamical behaviours.
In most multi-scale problems, such as the runaway electron dynamics in tokamaks, and the formation of energetic electrons in magnetosphere, long-term numerical integrations are required to reproduce the entire physical processes. For example, in tokamaks the typical timescale of runaway acceleration process is about  1s, which is $10^{8}$
times larger than its transit period. It is thus essential for the numerical algorithms to give
a correct, accurate, and fast long-term simulation in tracking the
secular particle trajectory. Conventional methods, such as  the
fourth order Runge-Kutta method, cannot trace the trajectory accurately after
a long time of computation due to the error accumulation. Great advances have been achieved in long-term accurate simulations of charged particle dynamics and Vlasov-Maxwell systems with the application of geometric integration methods \citep{Qin08-PRL,Qin09-PoP,Qin13-084503,Squire12,chin08-PRE,Finn05-pop,He15-JCP,He16-JCP,Zhang15-pop,Xiao13, Kraus14Phd,Zhou14,Shadwick14,Evstatiev13,Qin16SymVM,Xiao15Pop,He15KSym}.
Via preserving intrinsic structures of a dynamical system, geometric
integration methods \citep{Ruth83,Feng85,Feng10,Hairer03} usually
generate numerical results with better accuracy and conservative properties \citep{Hairer03,Shang99}.

The relativistic dynamics of a charged particle in the electromagnetic
fields ${\bf E}$ and ${\bf B}$ are governed by
\begin{equation}\label{eq:relLorentz}
\begin{aligned}
& \frac{d{\bf x}}{dt}=\frac{1}{m_{0}\gamma(p)}{\bf p},\\
 & \frac{d{\bf p}}{dt}=q{\bf E}({\bf x},t)+\frac{q}{m_{0}\gamma(p)}{\bf p}\times{\bf B}({\bf x},t),
\end{aligned}
\end{equation}
where ${\bf x}$ and ${\bf p}$ are the position and momentum vectors,
$m_{0}$ and $q$ denote the rest mass and charge of the particle,
and $\gamma(p)=\sqrt{1+p^{2}/(m_{0}^{2}c^{2})}$ is the Lorentz
factor with $c$ the speed of light in vacuum.
In Eq. (\ref{eq:relLorentz}), letting $p/c\rightarrow0$ leads to a non-relativistic
Lorentz force equation. Although the physical nature of the relativistic system is different
from the non-relativistic system, 
they have similar geometric properties, that is, the system (\ref{eq:relLorentz})
has the symplectic and volume-preserving properties \cite{He15-JCP,He15KSym}. It is believed that symplectic methods in general are implicit, and that popular explicit algorithms such as the Boris method is not symplectic \cite{Ellison2015489}.
Based on the volume-preserving property,  Symmetric Volume-Preserving (SVP) algorithms have been proposed
for solving the secular relativistic \cite{Zhang15-pop} and non-relativistic
\cite{Finn05-pop,He15-JCP,He16-JCP} dynamics of a charged particle.
One of the major properties of these methods is that the volume form in
phase space $({\bf x,p})$ is invariable along the updating map of
the numerical solution $\phi_{h}:({\bf x}_{k},{\bf p}_{k})\mapsto({\bf x}_{k+1},{\bf p}_{k+1})$,
which means the Jacobian
\[
\det\left(\frac{\partial({\bf x}_{k+1},{\bf p}_{k+1})}{\partial({\bf x}_{k},{\bf p}_{k})}\right)\equiv1.
\]
Another property is that the methods are time-symmetric, i.e. $\phi_h=\phi_{-h}^{-1}$.
The SVP methods have been verified to guarantee the long-term accuracy of numerical solutions and the
conservation of the constants of motion such as energy and angular
momentum. Moreover, they can be iterated explicitly and implemented easily, thus are efficient
in solving the secular trajectories of charged particles, and can
be developed as particle solvers in the Particle
In Cell (PIC) code \cite{Qiang04}.

In the current paper, we construct explicit, high order symmetric volume-preserving algorithms for
the relativistic dynamics under the general electromagnetic fields. Explicit volume-preserving algorithms can be constructed by the splitting technique \cite{fengs95vpa,Finn05-pop}. The equations are decomposed as a summation
of three incompressible subsystems, and SVP methods are constructed
by symmetric compositions of the  volume-preserving update mappings that solve the corresponding
subsystems. However, when the electromagnetic fields are time-dependent, it is not always trivial to solve the subsystems exactly.  
Therefore, we append the time $t$ to the dependent variables, it follows that the nonautonomous systems are turned into autonomous ones.  
In this case,  SVP methods can be given for general time-dependent electromagnetic fields by applying the splitting technique to the new system.  

As the SVP methods are developed using the splitting technique, it is known that the higher the order of accuracy is, the larger number of mappings is required in the compositions. This generates larger computing amount.
To reduce the computation amount over the simulation interval, 
we employ the processing technique \cite{Blanes99,Blanes06} in the construction of high order methods.
That is, we derive method in the form $\Psi_{h}=\chi_{h}\circ\Phi_{h}\circ\chi_{h}^{-1}$,
where the kernel $\Phi_{h}$   is the updating mapping
given by the usual splitting method,  the processor $\chi_{h}$  is a near identity map, and $\circ$ denotes the composition.
After $N$ steps of iteration, we have $\Psi_{h}^N=\chi_{h}\circ\Phi_{h}^N\circ\chi_{h}^{-1}$. From the relation it is easy to see that the computing efforts of $\Psi_{h}$ mainly comes from $\Phi_{h}$. A most efficient method can be derived by choosing the kernel method $\Phi_{h}$ as simple as possible,
and searching for the processor $\chi_{h}$  to achieve the desired
order of accuracy.  This idea has been applied to non-relativistic dynamical systems  \cite{He16-JCP}. For the relativistic dynamics, we split the motion equations in three parts or more, and present a high order SVP method by applying processing. We will show in the numerical experiments and the linear stability analysis that the newly derived high order methods are more efficient than the conventional composition methods, and allow larger step size to satisfy the stability  conditions.

This paper is organized as follows. In section 2, we give the derivation of
the SVP methods under the general time-dependent electromagnetic fields
using the splitting technique with processing. In section 3, we present the study of the linear stability of the
SVP methods. In section 4, the newly developed SVP methods are tested by two physical problems, i.e. the penning trap and the problem  possessing a plane polarized electromagnetic wave. 

\section{High order volume-preserving algorithms}

In this section, we give a general derivation of high order volume-preserving
algorithms for simulating the relativistic orbits under a time-dependent
electromagnetic field by using the splitting and processing technique.

We consider the most general case in which the electromagnetic fields
are time-dependent. 
To apply the  splitting and processing technique, we introduce $\sigma=t$ as a new depedent variable,
then it follows from (\ref{eq:relLorentz}) that
\begin{equation}\label{eq:tLorentz}
\begin{aligned}
{\frac{d}{dt}\left(\begin{array}{c}
{\bf x}\\
{\bf p}\\
\sigma
\end{array}\right)} & ={\left(\begin{array}{c}
\frac{1}{m_{0}\gamma(p)}{\bf p}\\
q{\bf E}({\bf x},\sigma)+\frac{q}{m_{0}\gamma(p)}{\bf p}\times{\bf B}({\bf x},\sigma)\\
1
\end{array}\right)}.
\end{aligned}
\end{equation}
From Eq.~(\ref{eq:tLorentz}), it is known that with the coordinate $({\bf x},{\bf p},\sigma)$
the system (\ref{eq:relLorentz}) becomes an autonomous system
defined in an expanded space $\mathbb{R}^{3}\times\mathbb{R}^{3}\times\mathbb{R}$
(see Ref.~\onlinecite{Blanes10} for more details). It is easy to check that the system (\ref{eq:tLorentz}) is source-free, i.e. the divergence of the vector field on the right hand side satisfies
$$\nabla_x\cdot \frac{1}{m_{0}\gamma(p)}{\bf p}+ \nabla_p\cdot \left(q{\bf E}({\bf x},\sigma)+\frac{q}{m_{0}\gamma(p)}{\bf p}\times{\bf B}({\bf x},\sigma)\right)+\nabla_{\sigma} 1 =0,$$
thus the volume in the expanded phase space is invariant along the exact solution flow.
Notice that for
any map $\Psi:({\bf x}_{k},{\bf p}_{k},\sigma_{k})\mapsto({\bf x}_{k+1},{\bf p}_{k+1},\sigma_{k+1})$
that preserves volume in the expanded space, the Jacobian satisfies
\[
1=\det\left(\frac{\partial({\bf x}_{k+1},{\bf p}_{k+1},\sigma_{k+1})}{\partial({\bf x}_{k},{\bf p}_{k},\sigma_{k})}\right)=\det\left(\frac{\partial({\bf x}_{k+1},{\bf p}_{k+1})}{\partial({\bf x}_{k},{\bf p}_{k})}\right),
\]
if $\frac{\partial\sigma_{k+1}}{\partial{\bf x}_{k}}=\frac{\partial\sigma_{k+1}}{\partial{\bf p}_{k}}={\bf 0}$,
$\frac{\partial\sigma_{k+1}}{\partial\sigma_{k}}=1$. 
This implies that if the appended variable $\sigma$ solves $\dot\sigma=const$, volume-preserving
methods for source-free systems in the expanded space also preserve
the volume of phase space $({\bf x},{\bf p})$. This gives us a hint on how to split the system.


Firstly, we split the system~(\ref{eq:tLorentz}). 
It is observed that the system~(\ref{eq:tLorentz}) can be decomposed as three source-free solvable subsystems,
\begin{equation}
\begin{aligned}{\frac{d}{dt}\left(\begin{array}{c}
{\bf x}\\
{\bf p}\\
\sigma
\end{array}\right)} & ={\left(\begin{array}{c}
\frac{1}{m_{0}\gamma(p)}{\bf p}\\
{\bf 0}\\
1
\end{array}\right)}+{\left(\begin{array}{c}
{\bf 0}\\
q{\bf E}({\bf x},\sigma)\\
0
\end{array}\right)}+{\left(\begin{array}{c}
{\bf 0}\\
\frac{q}{m_{0}\gamma(p)}{\bf p}\times{\bf B}({\bf x},\sigma)\\
0
\end{array}\right)}\\
 & =F_{1}({\bf x,p},t)+F_{2}({\bf x,p},t)+F_{3}({\bf x,p},t).
\end{aligned}
\label{eq:splitLorentz}
\end{equation}
The first two subsystems with $F_{1}$ and $F_{2}$ can be solved
exactly by a translation transformation as
\[
\phi_{h}^{F_{1}}:\left\{ \begin{array}{l}
{\bf x}(t+h)={\bf x}(t)+h\frac{{\bf p}(t)}{m_{0}\gamma(p(t))},\\
{\bf p}(t+h)={\bf p}(t),\\
\sigma(t+h)=\sigma(t)+h,
\end{array}\right.\quad\phi_{h}^{F_{2}}:\left\{ \begin{array}{l}
{\bf x}(t+h)={\bf x}(t),\\
{\bf p}(t+h)={\bf p}(t)+hq{\bf E}({\bf x}(t),\sigma(t)),\\
\sigma(t+h)=\sigma(t).
\end{array}\right.
\]
Here the mappings $\phi_{h}^{F_{i}}$, $i=1,2,3$ denote one $h$-time step updating
of the variables. When the third subsystem is concerned, it is noticed
that $p^{2}={\bf p}^{\top}{\bf p}$ is invariant along the exact solution
flow, so as to $\gamma(p)$. Thus, the updating map $\phi_{h}^{F_{3}}$
of the exact solution can be calculated as
\begin{subequations}\label{eq:3sub}
\begin{numcases}
{\phi_h^{F_3}:}
\mathbf{x}(t+h)=\mathbf{x}(t),\\
\mathbf{p}(t+h)=\exp\left(h\frac{q}{m_0\gamma(p(t))}\hat{\mathbf{B}}(\mathbf{x}(t),\sigma(t))\right)\mathbf{p}(t),\\
\sigma(t+h)=\sigma(t).
 \end{numcases}
\end{subequations}
with
$\hat{{\bf B}}=\left[\begin{array}{ccc}
0 & B_{3} & -B_{2}\\
-B_{3} & 0 & B_{1}\\
B_{2} & -B_{1} & 0
\end{array}\right]$ defined by ${\bf B}({\bf x})=[B_{1}({\bf x}),B_{2}({\bf x}),B_{3}({\bf x})]^{\top}$.
The operator $\exp$ in (\ref{eq:3sub}b) is the exponential operator
of a matrix, which can be expressed in a closed form for three dimensional
skew symmetric matrix as
\begin{equation}
\begin{aligned}
{\bf p}(t+h)&=\exp\left(ha\hat{{\bf B}}\right){\bf p}(t)\\
&={\bf p}(t)+\frac{\sin(haB)}{B}{\bf p}(t)\times{\bf B}+\frac{(1-\cos(haB))}{B^{2}}{\bf p}(t)\times{\bf B}\times{\bf B}.\end{aligned}
\label{eq:exp}
\end{equation}
Here $a=\frac{q}{m_0\gamma(p(t))}$.

It is easy to prove that each of the mappings $\varphi_{h}^{F_{1}}$,
$\varphi_{h}^{F_{2}}$, $\varphi_{h}^{F_{3}}$ preserves the volume in phase space $(\mathbf{x,p})$. Due to the group property, their various compositions
provide the SVP methods of any order \cite{Hairer03,Feng95vpa,Mclachlan02}.
As follows, we present some SVP methods of second and fourth orders.

\noindent \textbf{Second order symmetric methods}. A second order
symmetric method can be derived by the symmetric composition $G_{h}^{2}:=\phi_{\frac{h}{2}}^{F_{1}}\circ\phi_{\frac{h}{2}}^{F_{2}}\circ\phi_{h}^{F_{3}} \circ\phi_{\frac{h}{2}}^{F_{2}}\circ\phi_{\frac{h}{2}}^{F_{1}}$
\begin{align}
\begin{split}{\bf x}_{k+\frac{1}{2}} & ={\bf x}_{k}+\frac{h}{2}\frac{{\bf p}_{k}}{m_{0}\gamma(p_{k})},\\
{\bf p}^{-} & ={\bf p}_{k}+\frac{hq}{2}{\bf E}_{k+\frac{1}{2}},\\
{\bf p}^{+} & =\exp\left(\frac{hq}{m_{0}\gamma(p^{-})}\hat{{\bf B}}_{k+\frac{1}{2}}\right){\bf p}^{-},\\
{\bf p}_{k+1} & ={\bf p}^{+}+\frac{hq}{2}{\bf E}_{k+\frac{1}{2}},\\
{\bf x}_{k+1} & ={\bf x}_{k+\frac{1}{2}}+\frac{h}{2}\frac{{\bf p}_{k+1}}{m_{0}\gamma(p_{k+1})},
\end{split}
\label{vp:21}
\end{align}
where ${\bf E}_{k+\frac{1}{2}}:={\bf E}({\bf x}_{k+\frac{1}{2}},t_{k+\frac{1}{2}})$,
${\bf B}_{k+\frac{1}{2}}:={\bf B}({\bf x}_{k+\frac{1}{2}},t_{k+\frac{1}{2}})$
are the field values evaluated at the position ${\bf x}_{k+\frac{1}{2}}$
and the time $t_{k+\frac{1}{2}}$.

If we replace $\phi_{h}^{F_{3}}$ with a numerical solution $\Phi_h$ of the third subsystem, for example computed by the midpoint method, in symmetric
composition $G_{h}^{2}$, this provides an alternative SVP method of
second order
\begin{equation}\label{eq:tildeGh2}
\widetilde{G}_{h}^{2}:= \phi_{\frac{h}{2}}^{F_{1}}\circ\phi_{\frac{h}{2}}^{F_{2}}\circ \Phi_{h}^{F_{3}}\circ\phi_{\frac{h}{2}}^{F_{2}}\circ\phi_{\frac{h}{2}}^{F_{1}}.
\end{equation}
It recovers the numerical algorithm proposed in Ref.~\onlinecite{Zhang15-pop}.

In a similar way, the higher order SVP methods can be derived via various compositions
of approximate (exact) solutions of each subsystems. 
For example, the fourth order method can be derived by using the well
known Yoshida's composition \cite{Yoshida90} as
\begin{equation}
G_{h}^{4}Y=G_{a_{1}h}^{2}\circ G_{a_{2}h}^{2}\circ G_{a_{1}h}^{2},\label{eq:yoshida}
\end{equation}
or by using the Suzuki's fourth order composition as \cite{Suzuki92}
\begin{equation}
G_{h}^{4}S=G_{b_{1}h}^{2}\circ G_{b_{2}h}^{2}\circ G_{b_{3}h}^{2}\circ G_{b_{2}h}^{2}\circ G_{b_{1}h}^{2},\label{eq:suzuki}
\end{equation}
where $a_{1}=(2-2^{1/3})^{-1}$, $a_{2}=1-2a_{1}$, $b_{1}=b_{2}=(4-4^{1/3})^{-1}$,
$b_{3}=1-2(b_{1}+b_{2})$.  The
method $G_{h}^{4}S$ has smaller error constant than the method $G_{h}^{4}Y$.
It is clear from (\ref{eq:yoshida}) and (\ref{eq:suzuki}) that the
higher order methods produce the numerical solutions of high accuracy,
as well as the large computation cost. To reduce the computation cost,
we then present the efficient fourth order symmetric SVP methods
by employing the processing technique.

The main idea of processing technique is to apply a transformation
$\chi_{h}$ called the processor to a known lower order integrator
$\Phi_{h}$ such that the new derived method $\tilde{\Phi}_{h}=\chi_{h}\circ\Phi_{h}\circ\chi_{h}^{-1}$
has a higher order of accuracy than $\Phi_{h}$. Clearly, $\tilde{\Phi}_{h}$
maintains all properties (e.g. the long-term stability, structure-preserving
property) inherited by the lower order method $\Psi_{h}$. After $N$
steps of iteration it is $\tilde{\Psi}_{h}^{N}=\chi_{h}\circ\Psi_{h}^{N}\circ\chi_{h}^{-1}$
which states that using $\tilde{\Psi}_{h}^{N}$ does not need more
computation cost than $\Psi_{h}$. In Ref.~\onlinecite{He16-JCP},
processed methods are given when the system is separated into two
parts. For the relativistic dynamical system (\ref{eq:relLorentz})
with the splitting (\ref{eq:splitLorentz}), the kernel is given
by the compositions of   $G_{h}=\phi_{h}^{1}\circ\phi_{h}^{2}\circ\phi_{h}^{3}$
and $G_{h}^{*}=\phi_{h}^{3}\circ\phi_{h}^{2}\circ\phi_{h}^{1}$ as
\begin{equation}
\begin{aligned}\Psi_{h}=G_{a_{1}h}\circ G_{b_{1}h}^{*}\circ G_{a_{2}h}\circ G_{b_{2}h}^{*}\circ\ldots\circ G_{a_{s}h}\circ G_{b_{s}h}^{*},\\
\chi_{h}=G_{x_{1}h}\circ G_{y_{1}h}^{*}\circ G_{x_{2}h}\circ G_{y_{2}h}^{*}\circ\ldots\circ G_{x_{m}h}\circ G_{y_{m}h}^{*},
\end{aligned}
\label{eq:proceff}
\end{equation}
where $\{a_{i},b_{i}\}_{i=1}^{s}$ and $\{x_{i},y_{i}\}_{i=1}^{m}$
are the composition coefficients determined by the order conditions.
As an example, we list a fourth order processed method presented in Ref.\onlinecite{Blanes06}.

\noindent \textbf{Fourth order symmetric methods}.
One of processed composition methods reads
\begin{equation}
G_{h}^{4}P=\chi_{h}\circ\Psi_{h}\circ\chi_{h}^{-1},\label{eq:processed}
\end{equation}
where $\Psi_{h}$ and $\chi_{h}$ are in the form (\ref{eq:proceff})
with $s=m=4$, and the composition coefficients are listed in Table~\ref{tab:coeGh4P}.
It is easy to verify that the fourth order method $G_{h}^{4}P$ is symmetric, as $G_{h}^{4}P\circ G_{-h}^{4}P({\bf z})\equiv{\bf z}$
holds for any ${\bf z}$.
\begin{table}[h]
\centering %
\begin{tabular}{|ll|}
\hline
$a_{1}=\frac{\sqrt{18069}-15}{300}$  & $b_{1}=\frac{6}{25}$\tabularnewline
$a_{2}=\frac{9}{25}$  & $b_{2}=-\frac{\sqrt{18069}+15}{300}$ \tabularnewline
$a_{3}=b_{2}$, $a_{4}=b_{1}$  & $b_{3}=a_{2}$, $b_{4}=a_{1}$\tabularnewline
\hline
$x_{1}=0$  & $y_{1}=0.1171835753202670$\tabularnewline
$x_{2}=0.4731269439352653$  & $y_{2}=-0.1351671439946886$\tabularnewline
$x_{3}=1.350298160490375$  & $y_{3}=-0.4530449481299280$\tabularnewline
$x_{4}=0.05719279780976250$  & $y_{4}=-0.1930850894788554$\tabularnewline
\hline
\end{tabular}\caption{ Composition coefficients of the processed method $G_{h}^{4}P$.}
\label{tab:coeGh4P}
\end{table}

\section{Linear stability analysis}

The linear stability of the SVP methods applied to the non-relativistic
dynamics has been analyzed in Ref.~\onlinecite{He16-JCP}. In this section,
we generalize this study to relativistic dynamics. In order to do
this, we first present the test model equation. 

Consider the relativistic dynamics of a charged particle in an uniform
background magnetic field ${\bf B}=B_{0}\omega{\bf e}_{z}$, and electric
field produced by an ideal quadrupole potential distribution,
\[
\phi({\bf x})=\frac{1}{2}\frac{qB_{0}^{2}}{m_{0}}\epsilon(\lambda_{x}^{2}x^{2}+\lambda_{y}^{2}y^{2}-(\lambda_{x}^{2}+\lambda_{y}^{2})z^{2}),\quad\lambda_{x},\lambda_{y}>0,\epsilon=\pm1.
\]
Linearizing system (\ref{eq:relLorentz}) with the above electromagnetic field around $({\bf x}_{0},{\bf p}_{0})\in\mathbb{R}^{6}$,
we get the following equations
\begin{equation}
\dot{{\bf x}}=\frac{1}{\gamma_{0}}{\bf p},\quad\dot{{\bf p}}=-\frac{1}{B_{0}c}\nabla\phi({\bf x})+\frac{{\bf p}}{\gamma_{0}}\times\frac{{\bf B}}{B_{0}},\label{eq:sTest}
\end{equation}
where the variables are dimensionless normalized by $l_{0}=m_{0}c/(eB_{0})$
in space and $(\omega_{ce})^{-1}=m_{0}/(qB_{0})$ in time, and $\gamma_{0}=(\sqrt{1+p_{0}^{2}})^{3}>1$
is a constant. 
In the linearized system (\ref{eq:sTest}), the transverse motion and
the axial dynamics are decoupled. As the SVP methods developed in this
paper simulate this axial motion exactly, we only need to concentrate
on its transverse motion. Set $\lambda^{2}=\lambda_{x}^{2}=\lambda_{y}^{2}$,
and denote ${\bf x}=[x,y]$, ${\bf p}=[p_{x},p_{y}]$, the two-degree
test system is
\begin{equation}
\dot{{\bf x}}=\frac{1}{\gamma_{0}}{\bf p},\quad\dot{{\bf p}}=-\epsilon\lambda^{2}{\bf x}+\frac{\omega}{\gamma_{0}}J{\bf p},\label{eq:ssTest}
\end{equation}
where 
$J=\left(\begin{array}{cc}
0 & 1\\
-1 & 0
\end{array}\right)$ is the standard symplectic matrix.

Applying the SVP methods with time step $h$ to the test system (\ref{eq:ssTest}),
 we derive 
\begin{equation}
\left(\begin{array}{c}
\gamma_{0}{\bf x}^{k+1}\\
h{\bf p}^{k+1}
\end{array}\right)=M\left(\epsilon\left(\frac{h\lambda}{\sqrt{\gamma_{0}}}\right)^{2},\frac{h\omega}{\gamma_{0}}\right)\left(\begin{array}{c}
\gamma_{0}{\bf x}^{k}\\
h{\bf p}^{k}
\end{array}\right),\label{eq:stabMethod}
\end{equation}
where $M$ is the corresponding update matrix depending on $\epsilon(h\lambda/{\sqrt{\gamma_{0}}})^{2}$
and $h\omega/\gamma_{0}$. For the SVP methods constructed based on
the splitting method in Eq.(\ref{eq:splitLorentz}), the update matrix $M$ is the production
of update matrices for each subsystem. For the second order method $G_{h}=\phi_{h/2}^{F_{1}}\circ\phi_{h/2}^{F_{2}}\circ\phi_{h}^{F_{3}}\circ\phi_{h/2}^{F_{2}}\circ\phi_{h/2}^{F_{1}}$
in Eq. (\ref{vp:21}) applied to the test system (\ref{eq:ssTest}), $M$ is expressed as
\begin{equation}
M_{s}(h)=M_{s}^{1}\left(h/2\right)M_{s}^{2}\left(h/2\right)M_{s}^{3}(h)M_{s}^{2}\left(h/2\right)M_{s}^{1}\left(h/2\right),\label{eq:stab1}
\end{equation}
where $M_{s}^{1}(h/2)=\left[\begin{array}{cc}
I & h/2I\\
{\bf 0} & I
\end{array}\right],$ $M_{s}^{2}(h/2)=\left[\begin{array}{cc}
I & {\bf 0}\\
\,-\epsilon{h^{2}\lambda^{2}/(2\gamma_{0}})I & I
\end{array}\right],$ and $M_{s}^{3}(h)=\left[\begin{array}{cc}
I & {\bf 0}\\
{\bf 0} & O(h\omega/\gamma_{0})
\end{array}\right]$ are four-dimensional matrices, and $O(h\omega)$ is a rotation matrix
 $O(a)=\left[\begin{array}{cc}
\cos\left(a\right) & \sin\left(a\right)\\
-\sin\left(a\right) & \cos\left(a\right)
\end{array}\right].$ If replacing $h\omega$ with $2\arctan(h\omega/2)$ in Eq.(\ref{eq:stab1}),
we can get the evolution matrix of   the method $\widetilde{G}_{h}^{2}$ in Eq. (\ref{eq:tildeGh2}).

\begin{figure}[h]
\centering
{\includegraphics[width=6.5cm,height=5.3cm]{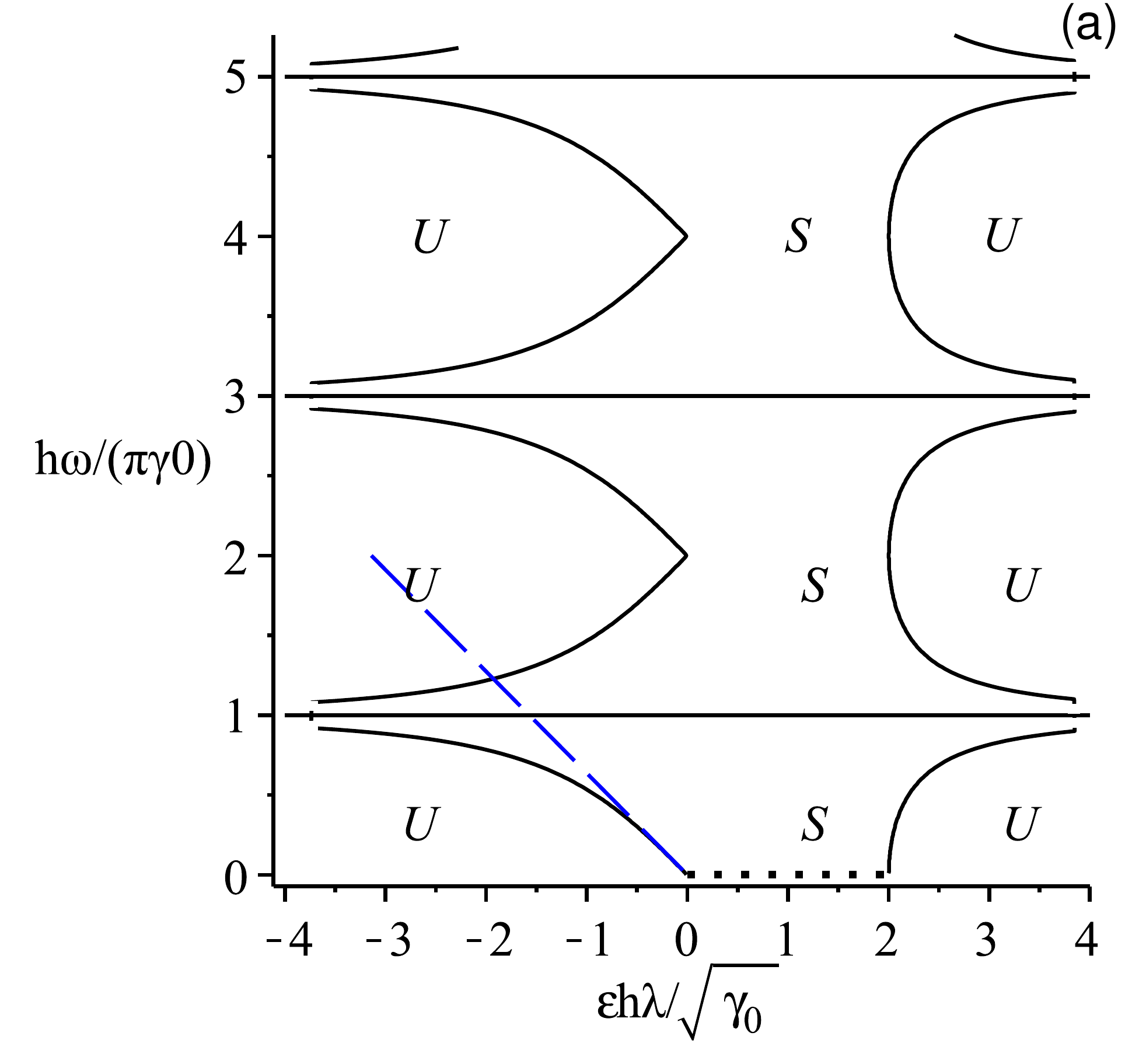}}
{\includegraphics[width=6.5cm,height=5.3cm]{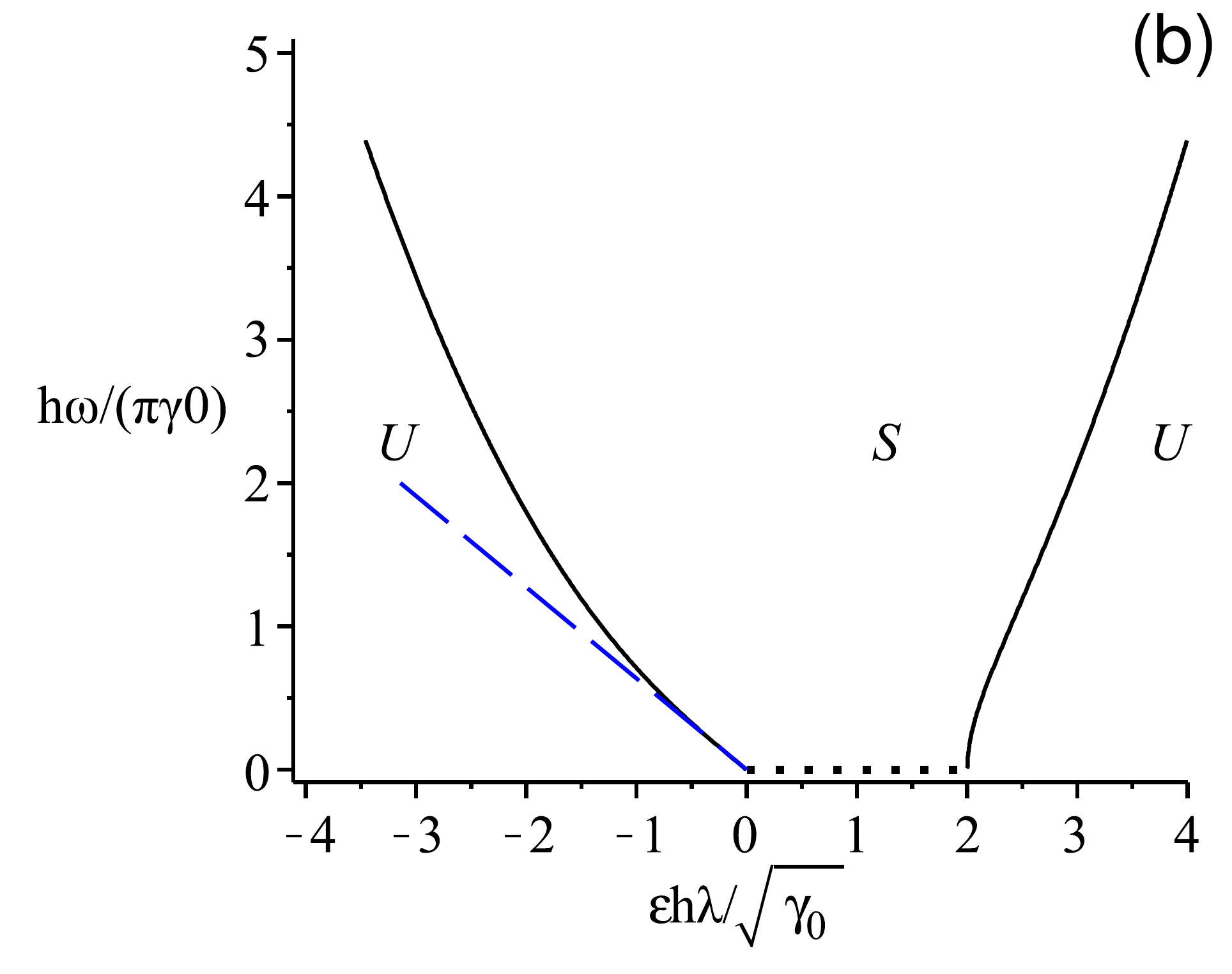}}
\caption{Stability domain of second order volume-preserving numerical methods.
Left(a): The method $G_{h}^{2}$ in Eq.(\ref{vp:21}); Right(b): The method $\widetilde{G}_{h}^{2}$ in Eq.(\ref{eq:tildeGh2}).
The abscissa represents $\epsilon h\lambda/\sqrt{\gamma_{0}}$, and
the ordinate represents $h\omega/(\gamma_{0}\pi)$. Here $\lambda^{2}$
reflects the dimensionless value ${\bf E}({\bf x}_{0})/B_{0}c$, $\omega$
reflects the dimensionless value $B({\bf x}_{0})/B_{0}$, and $\gamma_{0}=(\sqrt{1+p_{0}^{2}})^{3}$.
The solid curves are boundaries of the stability domain. `S' labels
stable region, `U' labels unstable region. The left bottom region
below the dashed line $\omega=-2\epsilon\lambda\sqrt{\gamma_{0}}$
is the unstable region of the test system.}
\label{fig:stabDomain}
\end{figure}

It is presented that a volume-preserving method applied to a source
free system is linearly stable if and only if the eigenvalues of the
update matrix have modulus 1 \cite{He16-JCP}. 
In Fig.~\ref{fig:stabDomain}, we display the stability domain of the
second order volume-preserving methods with respect to $\epsilon h\lambda/\sqrt{\gamma_{0}}$
and $h\omega/(\gamma_{0}\pi)$, where the left bottom region of the
blue dashed line indicates the physical unstable region of the test
system.

\begin{figure}[h]
\centering 
 {\includegraphics[width=7cm,height=3.8cm]{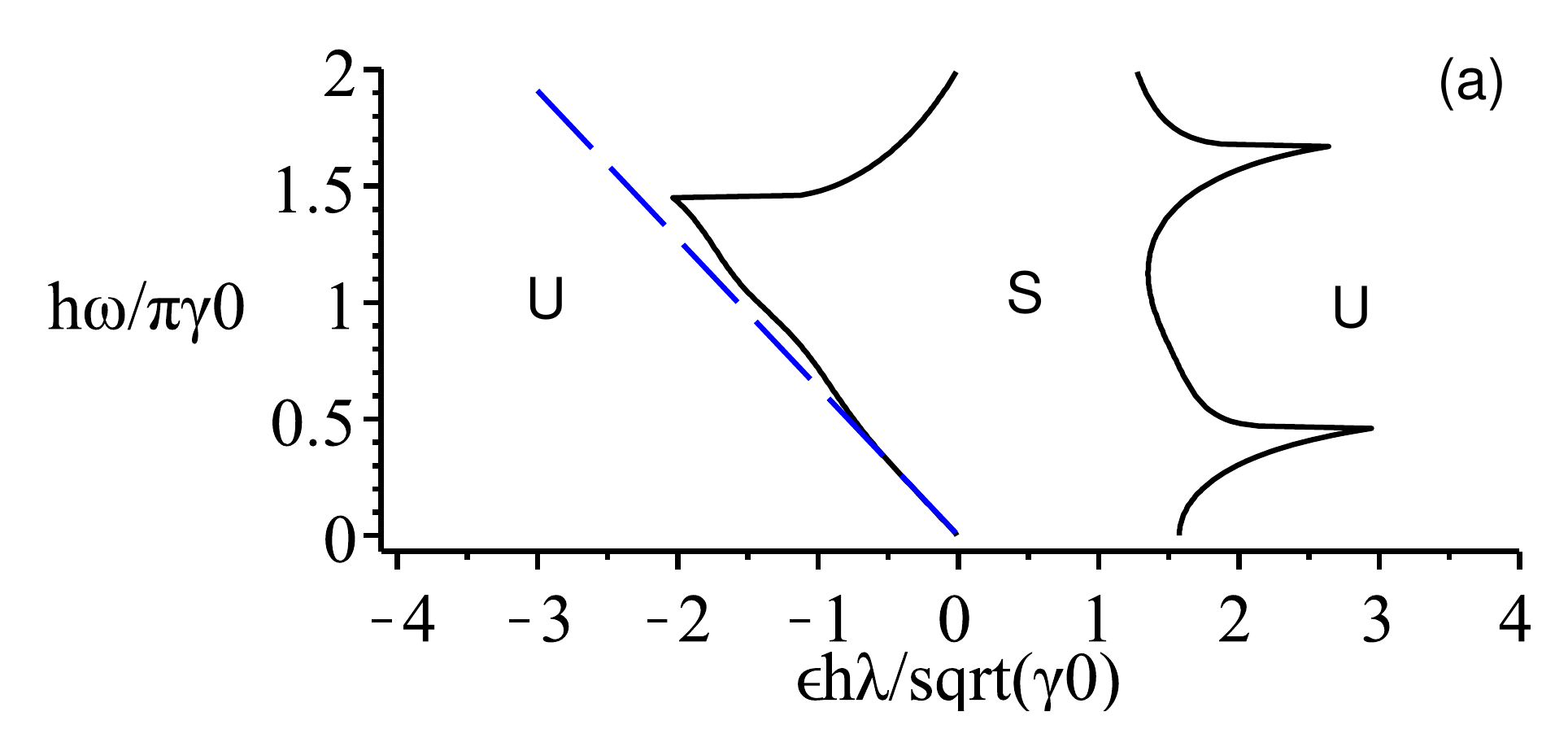}}
 {\includegraphics[width=7cm,height=3.8cm]{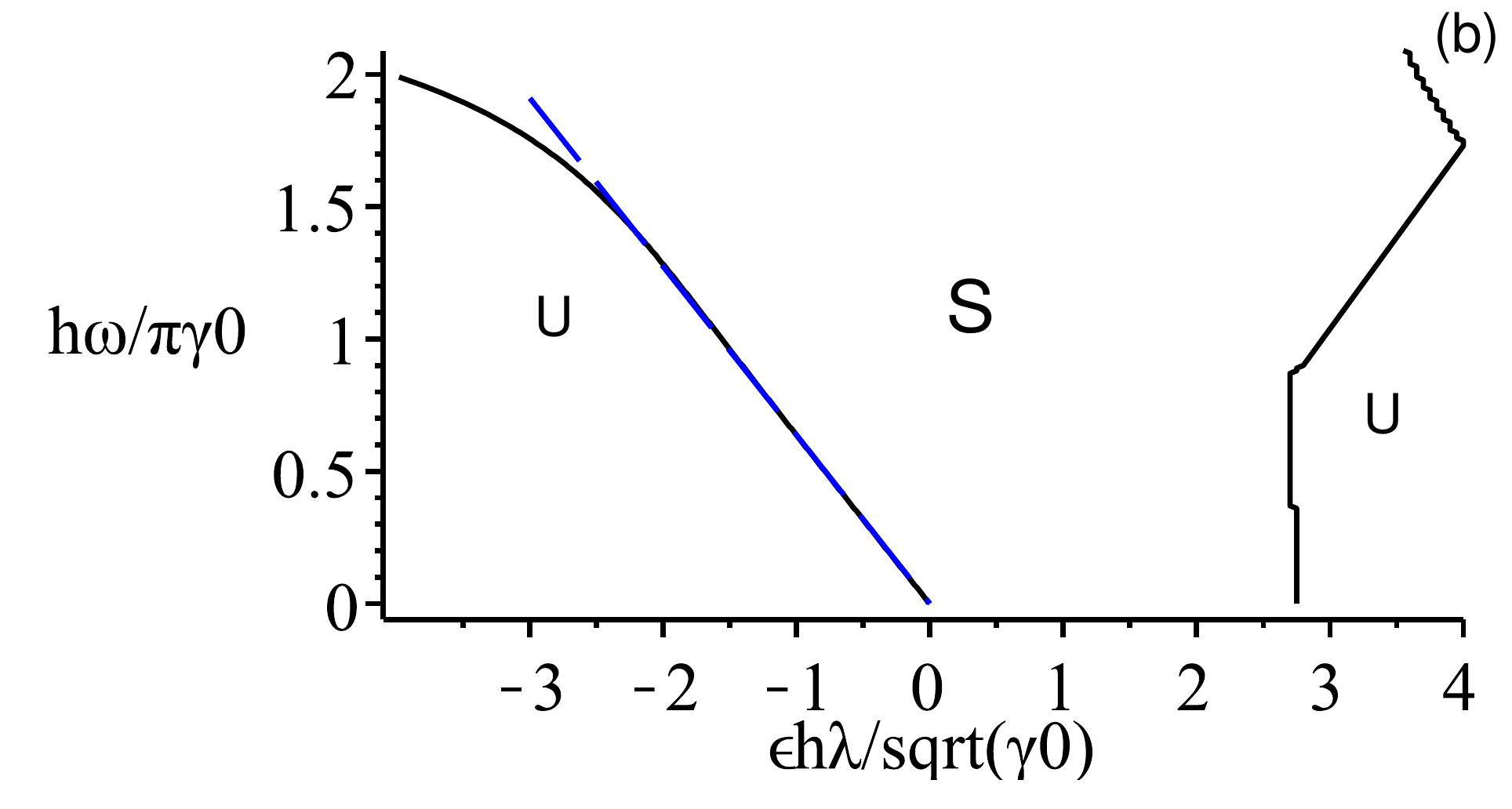}}\\
 {\includegraphics[width=9.5cm,height=3cm]{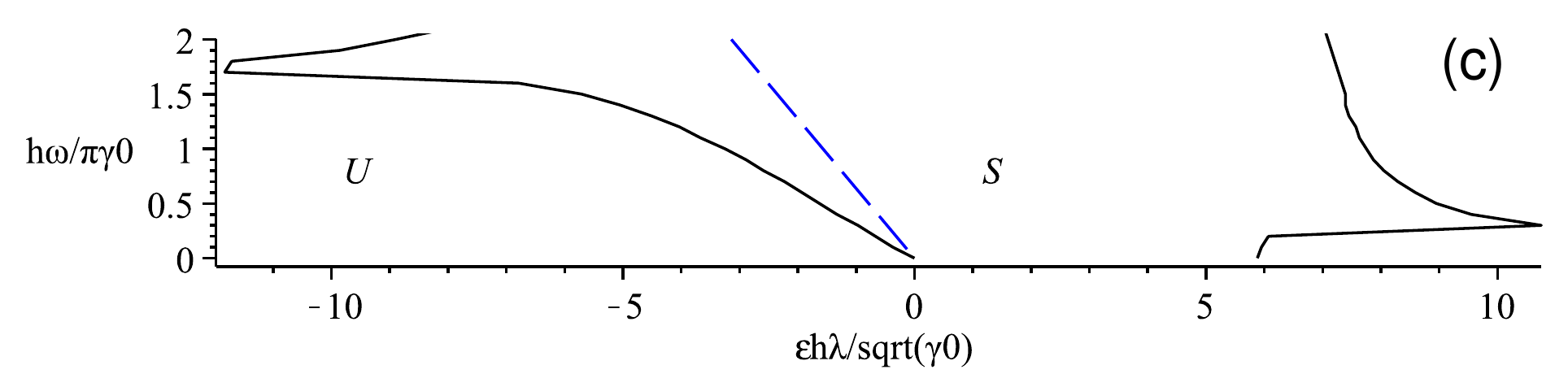}}
\caption{Stability domain of the fourth order volume-preserving methods. (a) The  Yoshida
composition based on the method $G_{h}^{2}$; (b). The   Suzuki composition
method based on $G_{h}^{2}$. (c). The processed fourth order method
$G_{h}^{4}P$. Here, $\lambda$, $\omega$ and $\gamma_{0}$ are defined
as above. }
\label{fig:stab4th} 
\end{figure}

From the observation of Fig.~\ref{fig:stabDomain}, we get the following
results:
\begin{enumerate}
\item When $\epsilon=1$, both of the two schemes are stable if $h\lambda/\sqrt{\gamma_{0}}<2$,
i.e., $h<2\sqrt{\gamma_{0}}/\lambda$. This means that if $\gamma_{0}$
is large or $\lambda$ is small, the larger $h$ can be taken
to guarantee that the second order SVP methods are still linearly stable. It is
noticed that larger $\gamma_{0}=(\sqrt{1+p_{0}^{2}})^{3}$ implies
larger kinetic energy.
\item If the two schemes $\widetilde{G}_{h}^{2}$ and $G_{h}^{2}$ are applied
to a problem with uniform electric field, i.e.,  $\lambda=0$, they are unconditionally stable. Moreover, for
the case when the electric field changes slowly in space, i.e. $\lambda$
is small enough the schemes $\widetilde{G}_{h}^{2}$ and $G_{h}^{2}$
can be stable for a very large $h$. In the practical computation
due to the Nyquist limit we need the time step $h$ satisfying $h\omega/(\gamma_{0}\pi)\leq1$
in order to simulate accurately the Larmor cyclotron.
\item From the two plots in Fig.\ref{fig:stabDomain}, it is observed that
the stability domain of the method $G_{h}^{2}$ is $2\pi$ periodic
with respect to $h\omega/\gamma_{0}$, while for the method $\widetilde{G}_{h}^{2}$
the domain becomes larger along with the increasing $h\omega/\gamma_{0}$
in $[0,5\pi]$. It is known that the slope of the line across the
origin of coordinate is $s=\omega/(\sqrt{\gamma_{0}}{\pi}\lambda)$.
With $\lambda$, $\omega$ and $\gamma_{0}$ satisfying $s<0.52$,
the stability domain shown in Fig.~\ref{fig:stabDomain} implies
that the method $G_{h}^{2}$ allows a larger time step than one for
the method ${\tilde{G}}_{h}^{2}$. 
\end{enumerate}
In Fig.~\ref{fig:stab4th}, the stability of the fourth order method
$G_{h}^{4}P$ in Eq.~(\ref{eq:processed}) are compared with the composed
methods $G_{h}^{4}Y$ in Eq.~(\ref{eq:yoshida}) and $G_{h}^{4}S$ in
Eq.~(\ref{eq:suzuki}). It is observed that compared with the second order
method $G_{h}^{2}$ in Fig.~\ref{fig:stabDomain}(a), the fourth
order Suzuki composition $G_{h}^{4}S$ has an enlarged stability domain
in Fig.~\ref{fig:stabDomain}(b), while the Yoshida composition has a shrunk stability domain
in Fig.~\ref{fig:stabDomain}(a). Among the three methods, the processed method $G_{h}^{4}P$
has the largest stability domain shown in (c). This verifies that
the processed method allows both higher order of accuracy and larger
threshold of the time step $h$.


\section{Numerical Experiments }

In this section, the SVP methods presented in the above section are
applied to simulate the relativistic problems with different electromagnetic fields. 

\textbf{Example 1.} Consider the relativistic dynamics of a charged
particle in the Penning trap. For this problem, the electromagnetic
field is given by
\begin{align*}
{\bf B} & =B_{0}{\bf e}_{z},\quad{\bf E}=-\epsilon E_{l}\left(\frac{x}{R_{0}}{\bf e}_{x}+\frac{y}{R_{0}}{\bf e}_{y}\right),
\end{align*}
where $B_{0}=1T$, $E_{l}=3V/m$, and $R_{0}=1m$.


We first simulate the relativistic dynamics of an electron in the
ideal penning trap with $\epsilon=1$. We take the initial momentum
as ${{\bf p}_{0}}_{\parallel}=0.1m_{0}c$, ${{\bf p}_{0}}_{\perp}=0.5m_{0}c$,
and the initial position as ${\bf x}=0.3l_{0}{\bf e}_{x}-l_{0}{\bf e}_{y}$.
After normalizing the temporal variables by $T_0=m_{0}/(eB_{0})=5.7\times10^{-12}s$, 
and the spatial variables by $l_{0}=m_{0}c/(eB_{0})$, the dimensionless
field parameters is ( $N$ denotes the normalized variable)
\[
{\bf B}_{N}=\omega{\bf e}_{z},{\bf E}_{N}=-\lambda^{2}\left(\frac{x}{R_{0}}{\bf e}_{x}+\frac{y}{R_{0}}{\bf e}_{y}\right),\text{~with~}\omega=1,\lambda=10^{-4}.
\]
In this experiment, as the initial kinetic energy is bounded and close
to $1$, we choose $\gamma_{0}=1$ in the test equation Eq.~(\ref{eq:ssTest}).
As $\epsilon=1$, and the slope $s=\omega/(\lambda\sqrt{\gamma_{0}}{\pi})=10^{4}/\pi$
is large enough, from Fig.~\ref{fig:stabDomain} we can see that
the two second order SVP methods $G_{h}^{2}$ and $\widetilde{G}_{h}^{2}$
are stable regardless of $h$. Thus the step size should be chosen
in $h\leq\gamma_{0}\pi/\omega=\pi$ according to the Nyquist limit.

\begin{figure}
\includegraphics[width=4cm,height=3.5cm]{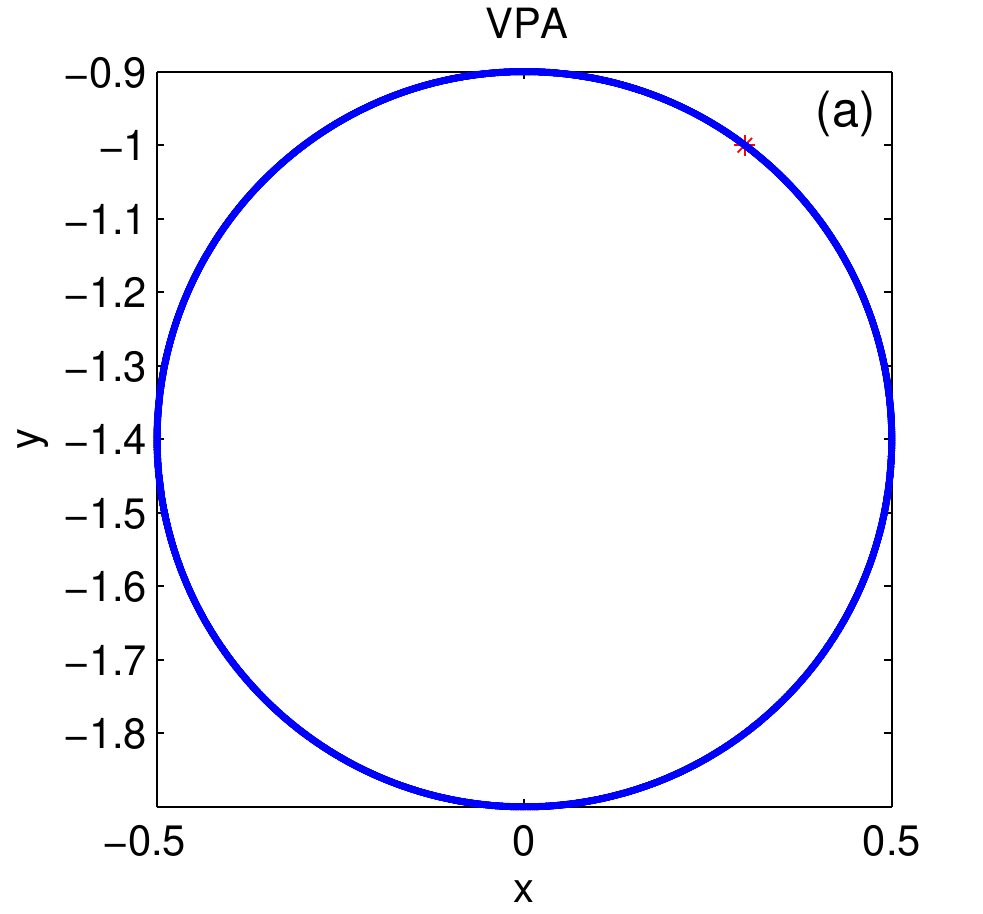}
{\includegraphics[width=4cm,height=3.5cm]{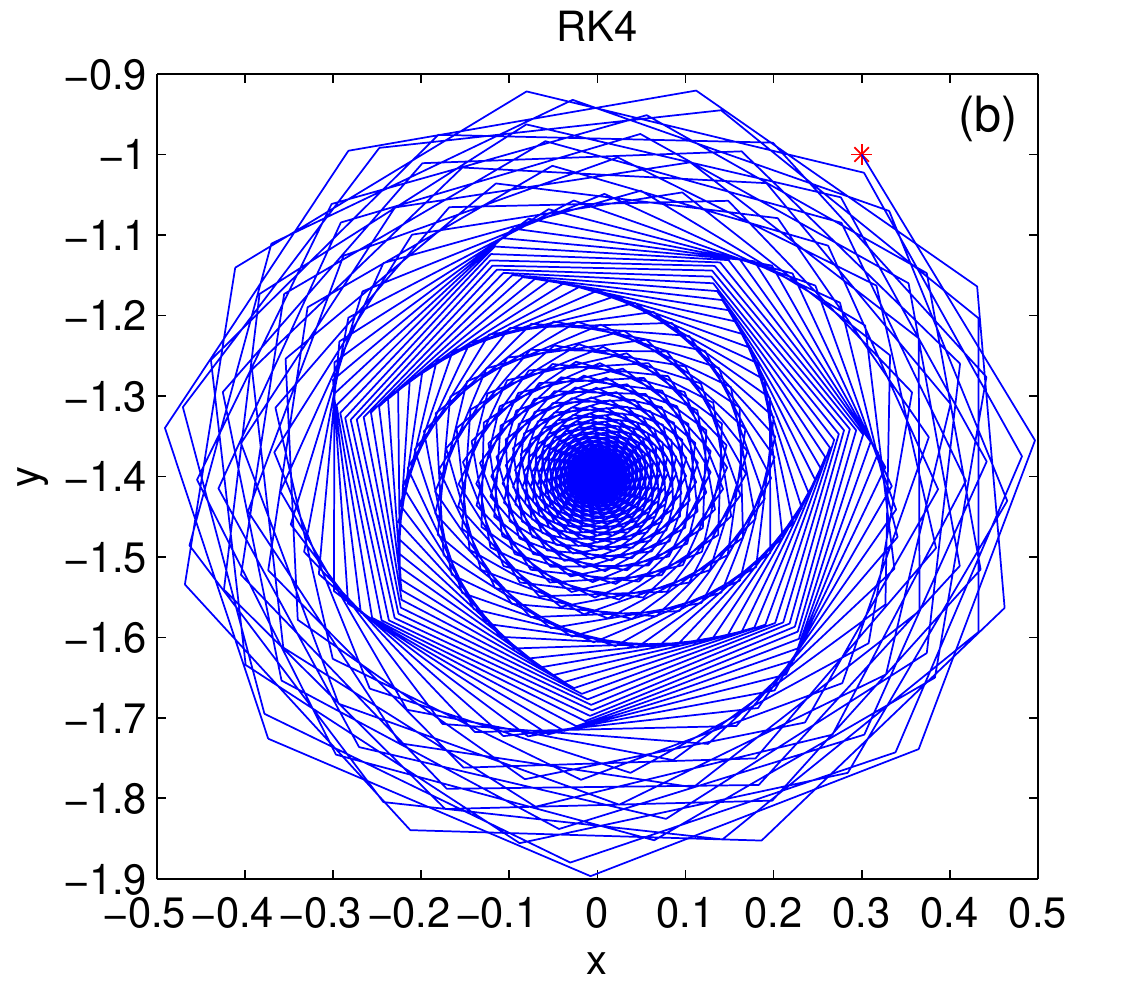}}\\
{\includegraphics[width=5cm,height=3cm]{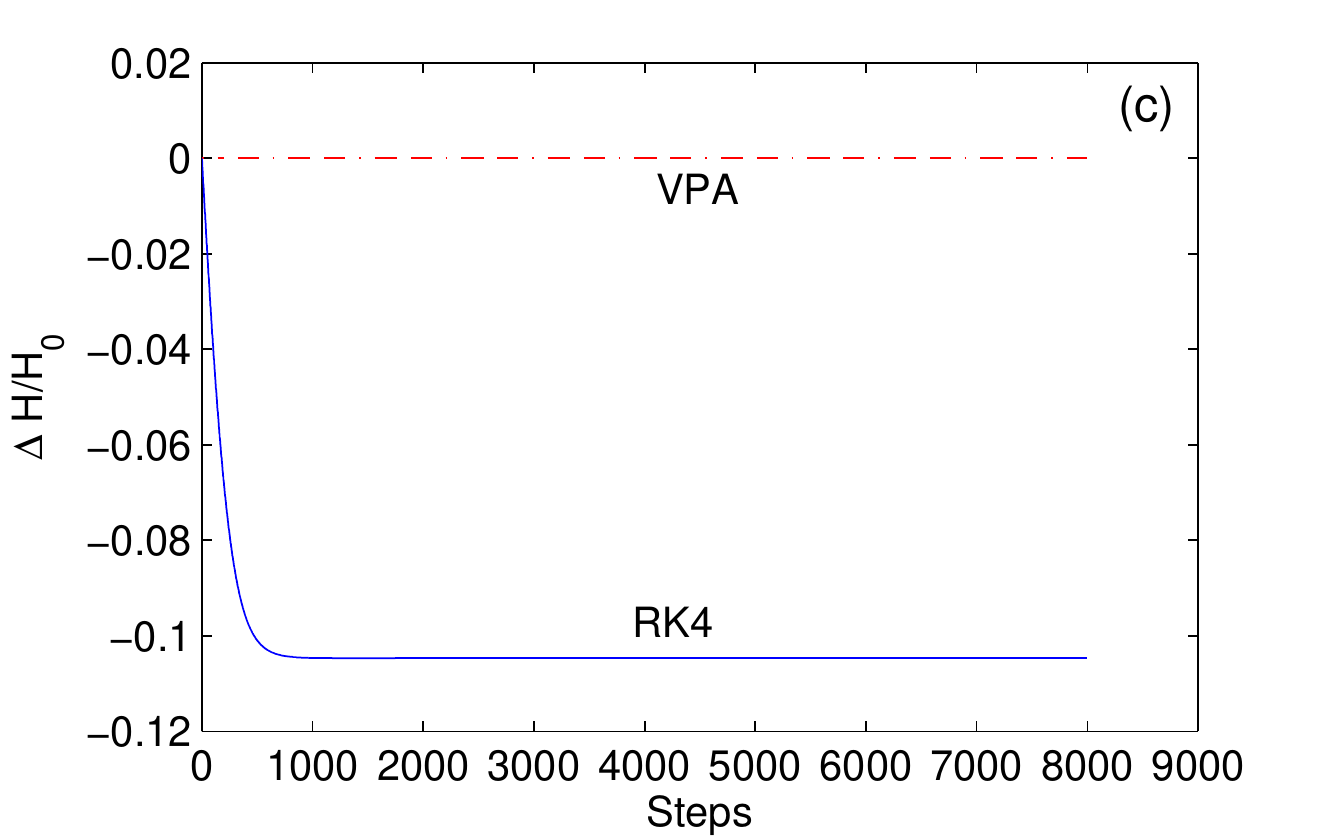}}
{\includegraphics[width=5cm,height=3cm]{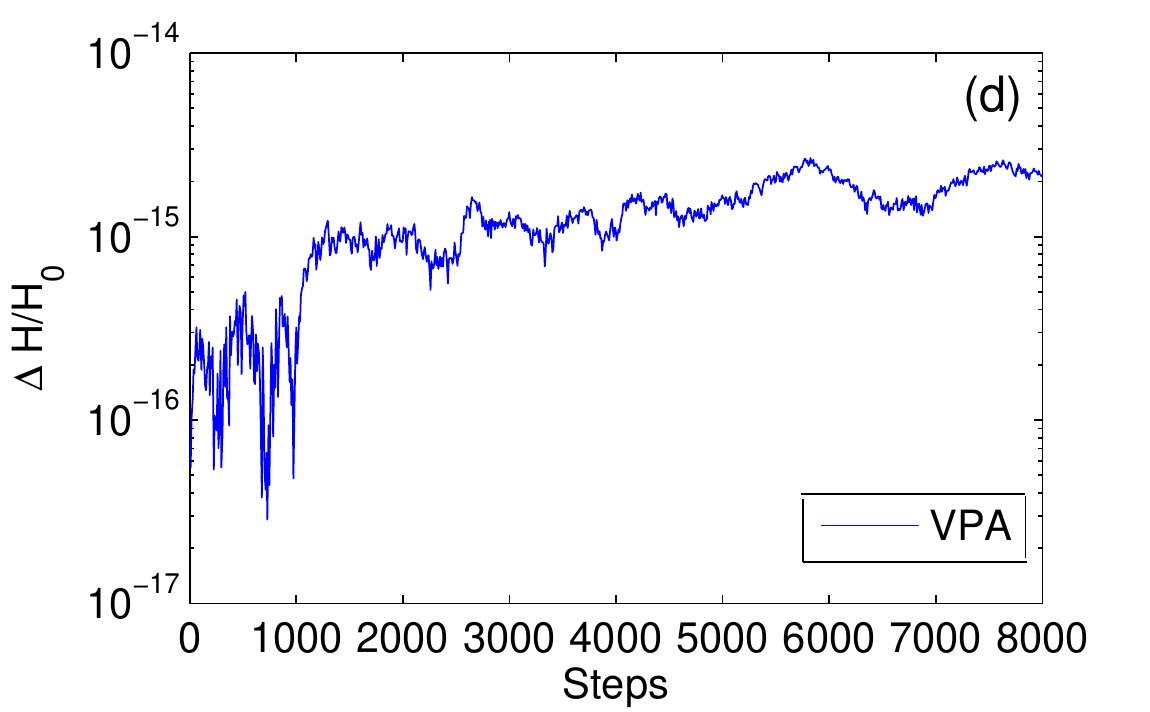}}
\caption{The simulation result of the relativistic dynamics of a particle in
an ideal penning trap.  The SVP methods and RK4 are iterated
for $8000$ steps with the step size $h=0.3\pi$. (a): Orbit by the SVP methods. (b): Orbit by  RK4. (c) and (d): Energy error as a function of steps.}
\label{fig:ex1a}
\end{figure}

In Fig.~\ref{fig:ex1a}, we show the numerical results computed by
the SVP methods running over $8000$ steps with $h=0.3\pi=5.37\times10^{-12}s$.
The explicit fourth order method RK4 is calculated as a comparison.
It is known that the exact orbit of the particle is an nearly
closed orbit with  radius $p_{\perp}/(m_{0}c)\approx0.5$.
It is observed from Fig.~\ref{fig:ex1a}(a) that the SVP method can simulate the orbit well.
The relative energy error displayed in Fig.~\ref{fig:ex1a}(b)
is bounded up to $10^{-14}$ during the entire simulation time. Conversely,
Fig.~\ref{fig:ex1a}(b) and (c) show that the numerical orbit spirals inside
and the energy error is damping.  This
is because of that  the numerical solution computed by RK4 scheme has the non-stability in long term computations.

\begin{figure}[h!]
\centering \subfigure{\includegraphics[width=5cm,height=3.5cm]{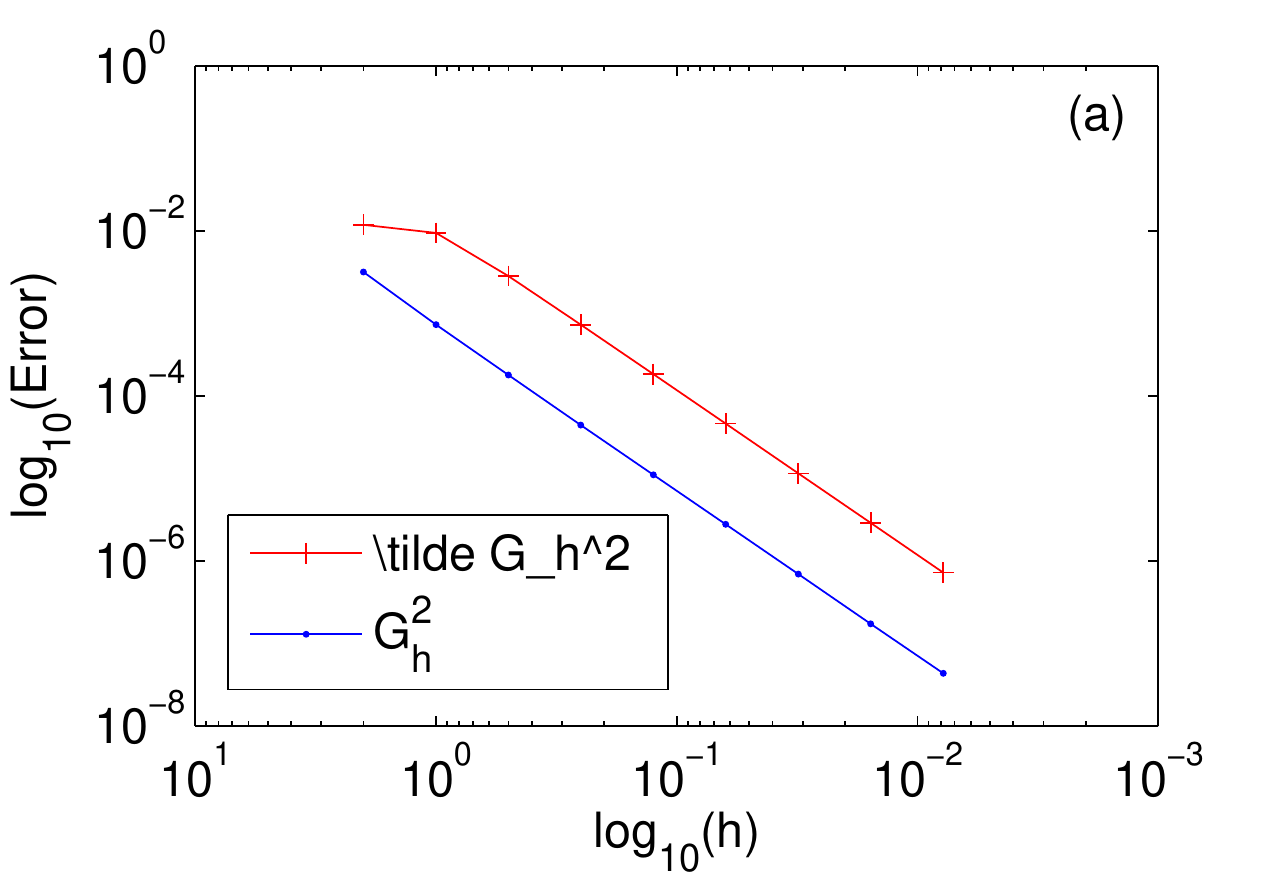}}
\subfigure{\includegraphics[width=5cm,height=3.5cm]{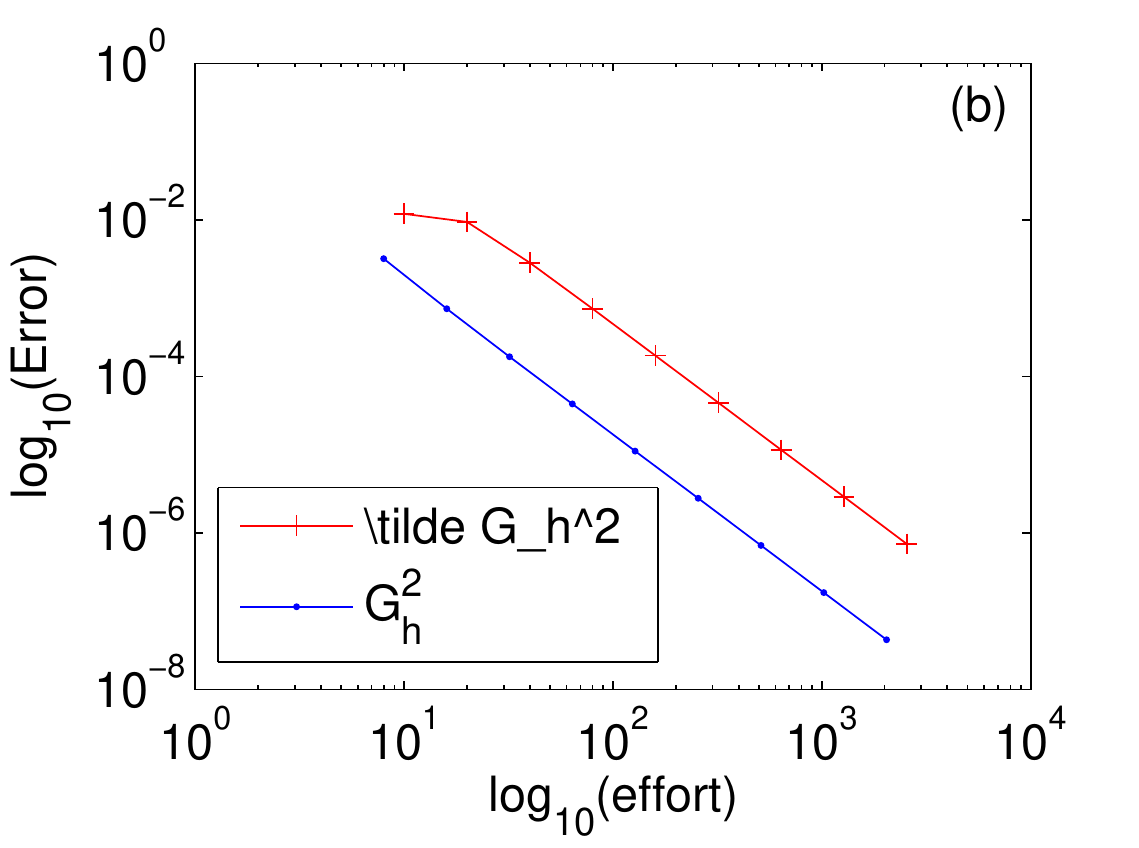}}\\
 \subfigure{\includegraphics[width=5cm,height=3.5cm]{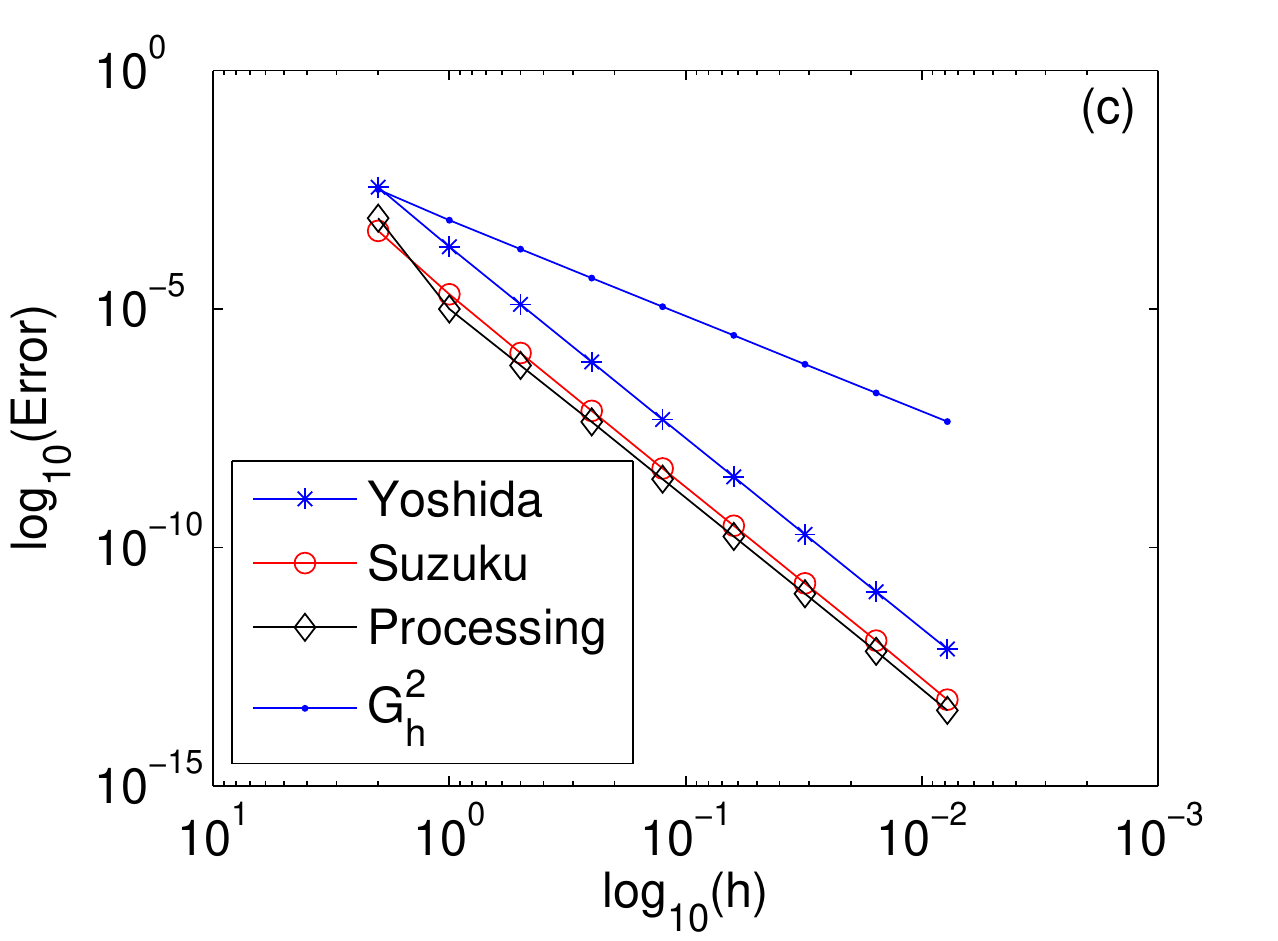}}
\subfigure{\includegraphics[width=5cm,height=3.5cm]{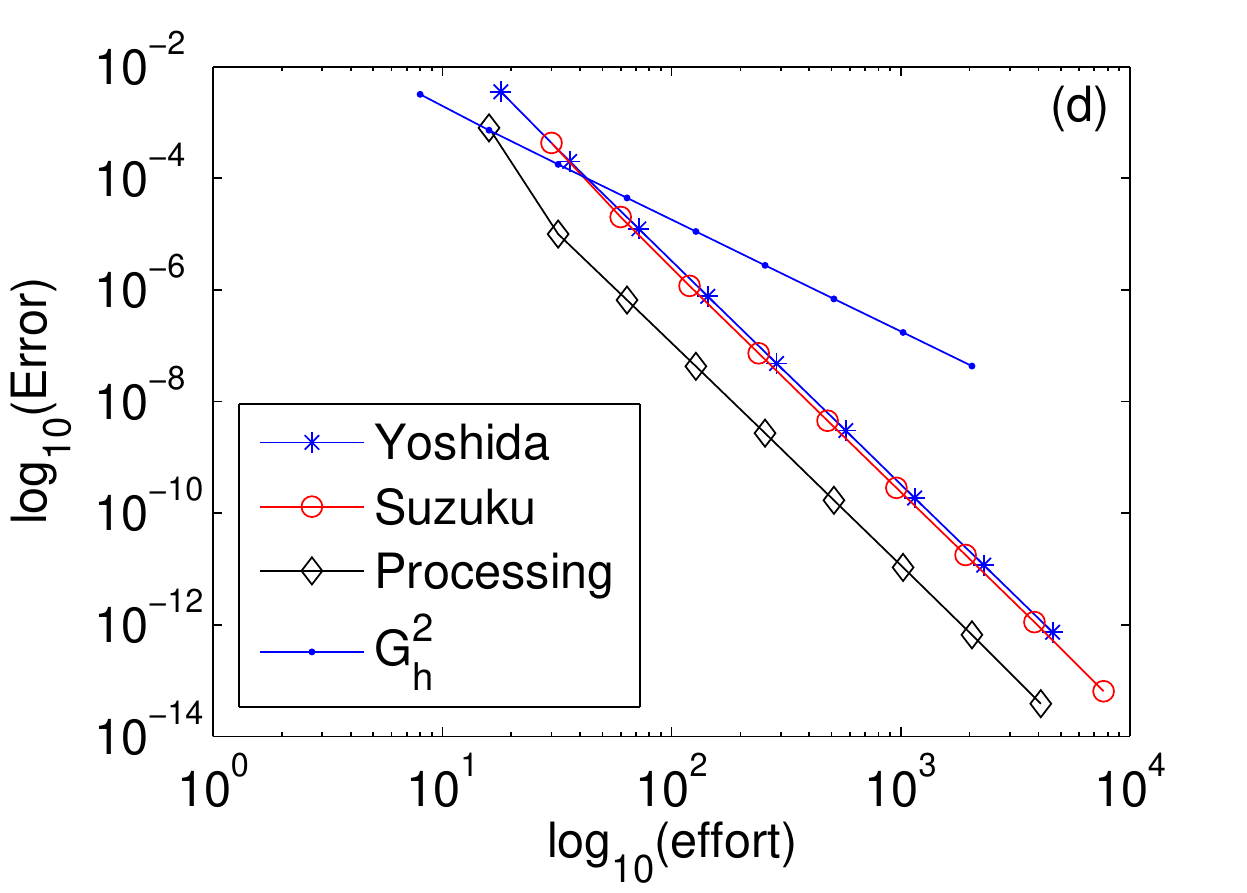}}
\caption{Relative errors of the dimensionless position variables in the experiment with an ideal penning trap. (a) and (c): Errors as a function of the time step $h$; (b) and (d): Errors as a function of the the computing amount (counted by evaluations of ${\bf E}$). (a) and (b): Errors of the 2nd order methods; (c) and (d): Errors of the 4-th order methods compared with second order method
$G_{h}^{2}$. }
\label{fig:ex1aCon}
\end{figure}

In Fig.~\ref{fig:ex1aCon}, the global errors of the dimensionless
position variables computed by the second and fourth order methods
are compared. Fig.~\ref{fig:ex1aCon}(a) and Fig.~\ref{fig:ex1aCon}(c) display the errors
as a function of time step $h$, which verifies the orders of the SVP
methods. 
In Fig.~\ref{fig:ex1aCon}(a), the method $G_{h}^{2}$ is more accurate than the method $\widetilde{G}_{h}^{2}$
because of the smaller error constant. In Fig.~\ref{fig:ex1aCon}(c), it is clear that the
processed fourth order method is the most accurate. Fig.~\ref{fig:ex1aCon}
(b) and Fig.~\ref{fig:ex1aCon}(d) display the errors as a function of the computing efforts,
which are counted by the number of the evaluations of ${\bf E}$.
It is observed that if the given tolerance on numerical errors is
less than $0.01\%$, the fourth order methods need less computing
efforts than the second order methods. Among the fourth order methods,
the processed method $G_{h}^{4}P$ is the cheapest.

Next we study the long-term performances of the SVP methods in the case with time-dependent electromagnetic fields.
The problem  possessing a plane polarized electromagnetic wave (see Ref.~\onlinecite{McMillan50}) is considered. After normalizing the variables as before, we choose the dimensionless fields to be $$\mathbf{E}=E_y\mathbf{e}_y, \mathbf{B}=B_z\mathbf{e}_z,E_y=B_z=3\sin(3(t-x)).$$
In this case, the evolution of the particle energy $W(\mathbf{p})$ satisfies
 $$I(t)=W(\mathbf{p}(t))-p_x(t)=constant,$$ where $p_x$ represents the momentum in the $x$-direction. Set the initial position and momentum to be $\mathbf{x}_0=0.3\mathbf{e}_x+0.2\mathbf{e}_y$, $\mathbf{p}_0=0.4\mathbf{e}_x+0.3\mathbf{e}_y+0.1\mathbf{e}_z$, we run the second order  SVP methods  for $10^6$ steps with the step size $h=0.1$. The fourth order Runge-Kutta method is  used as a comparison.
\begin{figure}[h!]
 \centering
{\includegraphics[width=7cm,height=5cm]{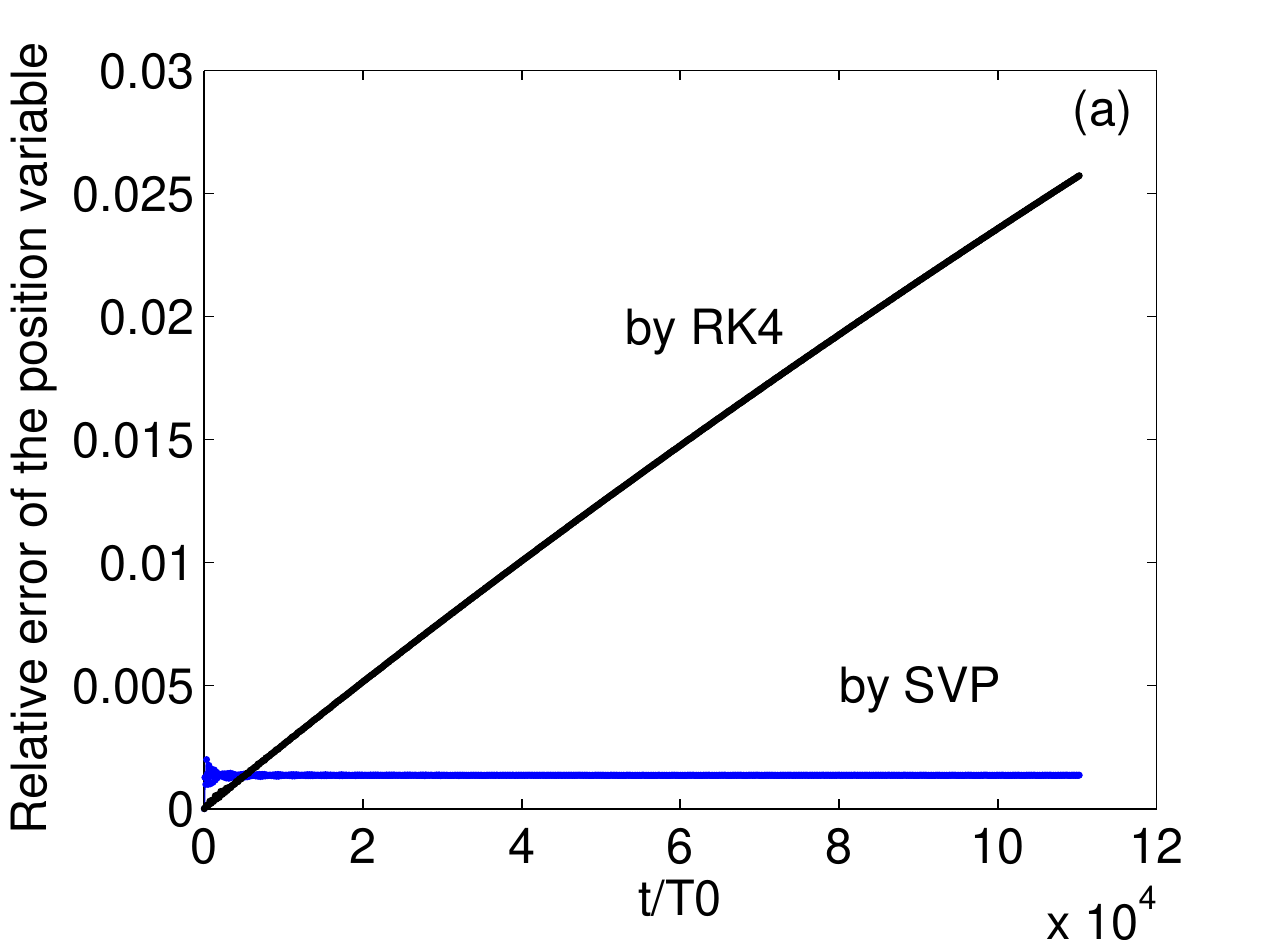}}
{\includegraphics[width=7cm,height=5cm]{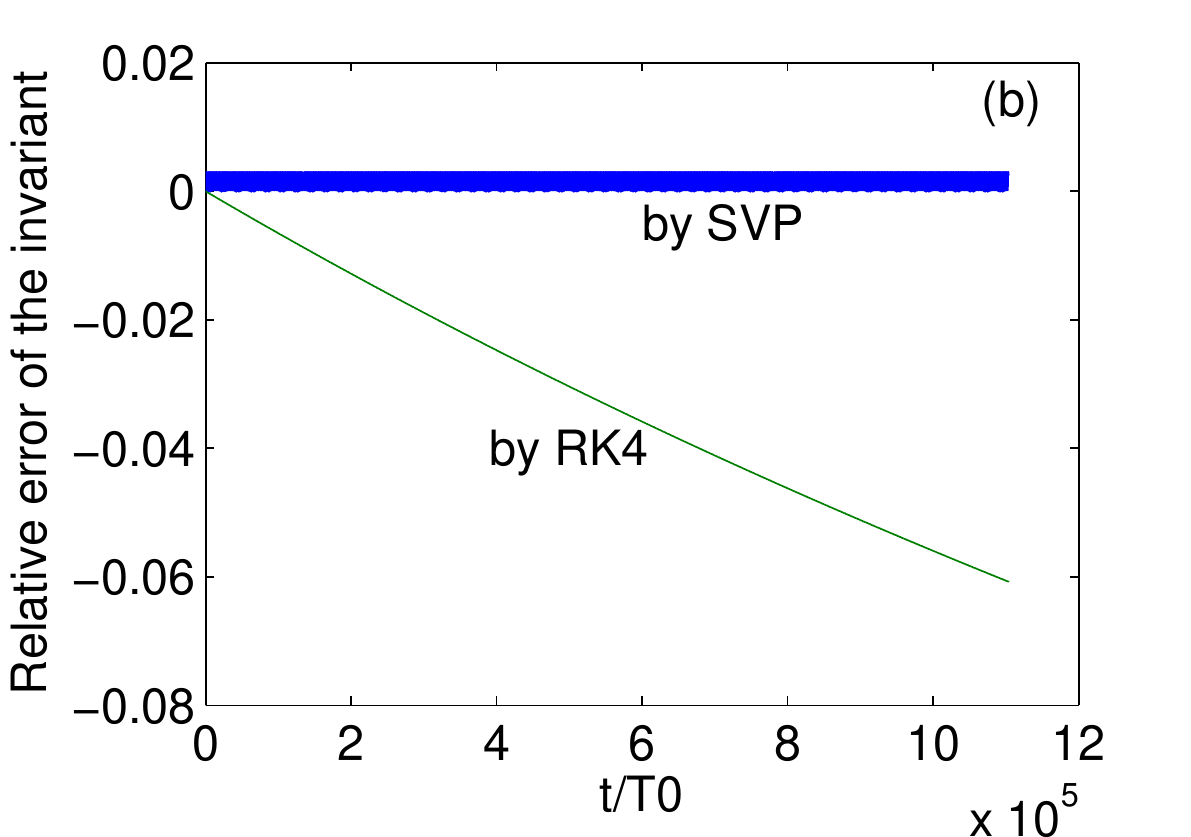}}
\caption{The long-term simulation result of the relativistic dynamics of a particle under a plane polarized electromagnetic wave. The step size is $h=0.1$. (a): Relative error of the position variables $\|\mathbf{x}_n-\mathbf{x}(nh)\|/\|\mathbf{x}(nh)\|$ as a function of normalized time $t/T_0$; (b): Realtive error of the invariant $I(t)$.}\label{fig:ex2}
\end{figure}
The results are shown in Fig.~\ref{fig:ex2}. From Fig.~\ref{fig:ex2}(a) we can see that the relative error of RK4 is smaller than that of the SVP method at the beginning few steps, but it grows over $1\%$ rapidly. Meanwhile the relative error of SVP methods stays below $0.5\%$ over the entire simulation time. In Fig.~\ref{fig:ex2}(b) the invariant $I(t)$ is preserved approximately by the SVP method, while the invariant  computed  by RK4 is dissipated. It  can be verified that for the long-term simulation  the two second order SVP methods are linearly and nonlinearly stable.

\section{Conclusion}

We have constructed high order volume-preserving methods for the relativistic
dynamics of a charged particle by the splitting technique with processing.
For the system with time-dependent fields, we reformulate  the system by extending its dependent variables space to include time $t$. With the newly derived system, we give a valid construction procedure of the
symmetric volume-preserving  methods. We have employed the processing technique to present the efficient methods with high order of accuracy.
 Linear stability which can serve as a hint on the choice
of time step size of the SVP methods are analyzed. Numerical experiments
show that the SVP methods are accurate and conservative for the long
term tracking of the trajectory of  relativistic particles.

\begin{acknowledgments}

This research was supported by ITER-China Program
(2015GB111003, 2014GB124005),
JSPS-NRF-NSFC A3 Foresight Program in the field of
Plasma Physics (NSFC-11261140328), the National Science
Foundation of China (11271357, 11575186, 11575185, 11505185, and
11505186), the Foundation for Innovative Research Groups of the NNSFC (11321061), the Fundamental Research Funds for
the Central Universities (WK2030040057).
\end{acknowledgments}

\bibliographystyle{apsrev4-1}
\bibliography{Refs}

\begin{thebibliography}{33}%
\makeatletter
\providecommand \@ifxundefined [1]{%
 \@ifx{#1\undefined}
}%
\providecommand \@ifnum [1]{%
 \ifnum #1\expandafter \@firstoftwo
 \else \expandafter \@secondoftwo
 \fi
}%
\providecommand \@ifx [1]{%
 \ifx #1\expandafter \@firstoftwo
 \else \expandafter \@secondoftwo
 \fi
}%
\providecommand \natexlab [1]{#1}%
\providecommand \enquote  [1]{``#1''}%
\providecommand \bibnamefont  [1]{#1}%
\providecommand \bibfnamefont [1]{#1}%
\providecommand \citenamefont [1]{#1}%
\providecommand \href@noop [0]{\@secondoftwo}%
\providecommand \href [0]{\begingroup \@sanitize@url \@href}%
\providecommand \@href[1]{\@@startlink{#1}\@@href}%
\providecommand \@@href[1]{\endgroup#1\@@endlink}%
\providecommand \@sanitize@url [0]{\catcode `\\12\catcode `\$12\catcode
  `\&12\catcode `\#12\catcode `\^12\catcode `\_12\catcode `\%12\relax}%
\providecommand \@@startlink[1]{}%
\providecommand \@@endlink[0]{}%
\providecommand \url  [0]{\begingroup\@sanitize@url \@url }%
\providecommand \@url [1]{\endgroup\@href {#1}{\urlprefix }}%
\providecommand \urlprefix  [0]{URL }%
\providecommand \Eprint [0]{\href }%
\providecommand \doibase [0]{http://dx.doi.org/}%
\providecommand \selectlanguage [0]{\@gobble}%
\providecommand \bibinfo  [0]{\@secondoftwo}%
\providecommand \bibfield  [0]{\@secondoftwo}%
\providecommand \translation [1]{[#1]}%
\providecommand \BibitemOpen [0]{}%
\providecommand \bibitemStop [0]{}%
\providecommand \bibitemNoStop [0]{.\EOS\space}%
\providecommand \EOS [0]{\spacefactor3000\relax}%
\providecommand \BibitemShut  [1]{\csname bibitem#1\endcsname}%
\let\auto@bib@innerbib\@empty
\bibitem [{\citenamefont {Qin}\ and\ \citenamefont {Guan}(2008)}]{Qin08-PRL}%
  \BibitemOpen
  \bibfield  {author} {\bibinfo {author} {\bibfnamefont {H.}~\bibnamefont
  {Qin}}\ and\ \bibinfo {author} {\bibfnamefont {X.}~\bibnamefont {Guan}},\
  }\href@noop {} {\bibfield  {journal} {\bibinfo  {journal} {Physical Review
  Letters}\ }\textbf {\bibinfo {volume} {100}},\ \bibinfo {pages} {035006}
  (\bibinfo {year} {2008})}\BibitemShut {NoStop}%
\bibitem [{\citenamefont {Qin}\ \emph {et~al.}(2009)\citenamefont {Qin},
  \citenamefont {Guan},\ and\ \citenamefont {Tang}}]{Qin09-PoP}%
  \BibitemOpen
  \bibfield  {author} {\bibinfo {author} {\bibfnamefont {H.}~\bibnamefont
  {Qin}}, \bibinfo {author} {\bibfnamefont {X.}~\bibnamefont {Guan}}, \ and\
  \bibinfo {author} {\bibfnamefont {W.~M.}\ \bibnamefont {Tang}},\ }\href@noop
  {} {\bibfield  {journal} {\bibinfo  {journal} {Physics of Plasmas}\ }\textbf
  {\bibinfo {volume} {16}},\ \bibinfo {pages} {042510} (\bibinfo {year}
  {2009})}\BibitemShut {NoStop}%
\bibitem [{\citenamefont {Qin}\ \emph {et~al.}(2013)\citenamefont {Qin},
  \citenamefont {Zhang}, \citenamefont {Xiao}, \citenamefont {Liu},
  \citenamefont {Sun},\ and\ \citenamefont {Tang}}]{Qin13-084503}%
  \BibitemOpen
  \bibfield  {author} {\bibinfo {author} {\bibfnamefont {H.}~\bibnamefont
  {Qin}}, \bibinfo {author} {\bibfnamefont {S.}~\bibnamefont {Zhang}}, \bibinfo
  {author} {\bibfnamefont {J.}~\bibnamefont {Xiao}}, \bibinfo {author}
  {\bibfnamefont {J.}~\bibnamefont {Liu}}, \bibinfo {author} {\bibfnamefont
  {Y.}~\bibnamefont {Sun}}, \ and\ \bibinfo {author} {\bibfnamefont {W.~M.}\
  \bibnamefont {Tang}},\ }\href@noop {} {\bibfield  {journal} {\bibinfo
  {journal} {Physics of Plasmas}\ }\textbf {\bibinfo {volume} {20}},\ \bibinfo
  {pages} {084503} (\bibinfo {year} {2013})}\BibitemShut {NoStop}%
\bibitem [{\citenamefont {Squire}\ \emph {et~al.}(2012)\citenamefont {Squire},
  \citenamefont {Qin},\ and\ \citenamefont {Tang}}]{Squire12}%
  \BibitemOpen
  \bibfield  {author} {\bibinfo {author} {\bibfnamefont {J.}~\bibnamefont
  {Squire}}, \bibinfo {author} {\bibfnamefont {H.}~\bibnamefont {Qin}}, \ and\
  \bibinfo {author} {\bibfnamefont {W.~M.}\ \bibnamefont {Tang}},\ }\href@noop
  {} {\bibfield  {journal} {\bibinfo  {journal} {Physics of Plasmas}\ }\textbf
  {\bibinfo {volume} {19}},\ \bibinfo {pages} {084501} (\bibinfo {year}
  {2012})}\BibitemShut {NoStop}%
\bibitem [{\citenamefont {Chin}(2008)}]{chin08-PRE}%
  \BibitemOpen
  \bibfield  {author} {\bibinfo {author} {\bibfnamefont {S.~A.}\ \bibnamefont
  {Chin}},\ }\href@noop {} {\bibfield  {journal} {\bibinfo  {journal} {Phy.
  Rev. E}\ }\textbf {\bibinfo {volume} {77}},\ \bibinfo {pages} {066401}
  (\bibinfo {year} {2008})}\BibitemShut {NoStop}%
\bibitem [{\citenamefont {Finn}\ and\ \citenamefont
  {Chac\'{o}n}(2005)}]{Finn05-pop}%
  \BibitemOpen
  \bibfield  {author} {\bibinfo {author} {\bibfnamefont {J.~M.}\ \bibnamefont
  {Finn}}\ and\ \bibinfo {author} {\bibfnamefont {L.}~\bibnamefont
  {Chac\'{o}n}},\ }\href
  {http://scitation.aip.org/content/aip/journal/pop/12/5/10.1063/1.1889156}
  {\bibfield  {journal} {\bibinfo  {journal} {Physics of Plasmas}\ }\textbf
  {\bibinfo {volume} {12}},\ \bibinfo {eid} {054503} (\bibinfo {year}
  {2005})}\BibitemShut {NoStop}%
\bibitem [{\citenamefont {He}\ \emph {et~al.}(2015{\natexlab{a}})\citenamefont
  {He}, \citenamefont {Sun}, \citenamefont {Liu},\ and\ \citenamefont
  {Qin}}]{He15-JCP}%
  \BibitemOpen
  \bibfield  {author} {\bibinfo {author} {\bibfnamefont {Y.}~\bibnamefont
  {He}}, \bibinfo {author} {\bibfnamefont {Y.}~\bibnamefont {Sun}}, \bibinfo
  {author} {\bibfnamefont {J.}~\bibnamefont {Liu}}, \ and\ \bibinfo {author}
  {\bibfnamefont {H.}~\bibnamefont {Qin}},\ }\href@noop {} {\bibfield
  {journal} {\bibinfo  {journal} {Journal of Computational Physics}\ }\textbf
  {\bibinfo {volume} {281}},\ \bibinfo {pages} {135} (\bibinfo {year}
  {2015}{\natexlab{a}})}\BibitemShut {NoStop}%
\bibitem [{\citenamefont {He}\ \emph {et~al.}(2016)\citenamefont {He},
  \citenamefont {Sun}, \citenamefont {Liu},\ and\ \citenamefont
  {Qin}}]{He16-JCP}%
  \BibitemOpen
  \bibfield  {author} {\bibinfo {author} {\bibfnamefont {Y.}~\bibnamefont
  {He}}, \bibinfo {author} {\bibfnamefont {Y.}~\bibnamefont {Sun}}, \bibinfo
  {author} {\bibfnamefont {J.}~\bibnamefont {Liu}}, \ and\ \bibinfo {author}
  {\bibfnamefont {H.}~\bibnamefont {Qin}},\ }\href@noop {} {\bibfield
  {journal} {\bibinfo  {journal} {Journal of Computational Physics}\ }\textbf
  {\bibinfo {volume} {305}},\ \bibinfo {pages} {172} (\bibinfo {year}
  {2016})}\BibitemShut {NoStop}%
\bibitem [{\citenamefont {Zhang}\ \emph {et~al.}(2015)\citenamefont {Zhang},
  \citenamefont {Liu}, \citenamefont {Qin}, \citenamefont {Wang}, \citenamefont
  {He},\ and\ \citenamefont {Sun}}]{Zhang15-pop}%
  \BibitemOpen
  \bibfield  {author} {\bibinfo {author} {\bibfnamefont {R.}~\bibnamefont
  {Zhang}}, \bibinfo {author} {\bibfnamefont {J.}~\bibnamefont {Liu}}, \bibinfo
  {author} {\bibfnamefont {H.}~\bibnamefont {Qin}}, \bibinfo {author}
  {\bibfnamefont {Y.}~\bibnamefont {Wang}}, \bibinfo {author} {\bibfnamefont
  {Y.}~\bibnamefont {He}}, \ and\ \bibinfo {author} {\bibfnamefont
  {Y.}~\bibnamefont {Sun}},\ }\href@noop {} {\bibfield  {journal} {\bibinfo
  {journal} {Physics of Plasmas}\ ,\ \bibinfo {pages} {Submitted}} (\bibinfo
  {year} {2015})}\BibitemShut {NoStop}%
\bibitem [{\citenamefont {Xiao}\ \emph {et~al.}(2013)\citenamefont {Xiao},
  \citenamefont {Liu}, \citenamefont {Qin},\ and\ \citenamefont {Yu}}]{Xiao13}%
  \BibitemOpen
  \bibfield  {author} {\bibinfo {author} {\bibfnamefont {J.}~\bibnamefont
  {Xiao}}, \bibinfo {author} {\bibfnamefont {J.}~\bibnamefont {Liu}}, \bibinfo
  {author} {\bibfnamefont {H.}~\bibnamefont {Qin}}, \ and\ \bibinfo {author}
  {\bibfnamefont {Z.}~\bibnamefont {Yu}},\ }\href@noop {} {\bibfield  {journal}
  {\bibinfo  {journal} {Physics of Plasmas}\ }\textbf {\bibinfo {volume}
  {20}},\ \bibinfo {pages} {102517} (\bibinfo {year} {2013})}\BibitemShut
  {NoStop}%
\bibitem [{\citenamefont {Kraus}(2014)}]{Kraus14Phd}%
  \BibitemOpen
  \bibfield  {author} {\bibinfo {author} {\bibfnamefont {M.}~\bibnamefont
  {Kraus}},\ }\emph {\bibinfo {title} {Variational Integrators in Plasma
  Physics}},\ \href@noop {} {Ph.D. thesis},\ \bibinfo  {school} {Technical
  University of Munich} (\bibinfo {year} {2014})\BibitemShut {NoStop}%
\bibitem [{\citenamefont {Zhou}\ \emph {et~al.}(2014)\citenamefont {Zhou},
  \citenamefont {Qin}, \citenamefont {Burby},\ and\ \citenamefont
  {Bhattacharjee}}]{Zhou14}%
  \BibitemOpen
  \bibfield  {author} {\bibinfo {author} {\bibfnamefont {Y.}~\bibnamefont
  {Zhou}}, \bibinfo {author} {\bibfnamefont {H.}~\bibnamefont {Qin}}, \bibinfo
  {author} {\bibfnamefont {J.~W.}\ \bibnamefont {Burby}}, \ and\ \bibinfo
  {author} {\bibfnamefont {A.}~\bibnamefont {Bhattacharjee}},\ }\href@noop {}
  {\bibfield  {journal} {\bibinfo  {journal} {Physics of Plasmas}\ }\textbf
  {\bibinfo {volume} {21}},\ \bibinfo {pages} {102109} (\bibinfo {year}
  {2014})}\BibitemShut {NoStop}%
\bibitem [{\citenamefont {Shadwick}\ \emph {et~al.}(2014)\citenamefont
  {Shadwick}, \citenamefont {Stamm},\ and\ \citenamefont
  {Evstatiev}}]{Shadwick14}%
  \BibitemOpen
  \bibfield  {author} {\bibinfo {author} {\bibfnamefont {B.~A.}\ \bibnamefont
  {Shadwick}}, \bibinfo {author} {\bibfnamefont {A.~B.}\ \bibnamefont {Stamm}},
  \ and\ \bibinfo {author} {\bibfnamefont {E.~G.}\ \bibnamefont {Evstatiev}},\
  }\href@noop {} {\bibfield  {journal} {\bibinfo  {journal} {Physics of
  Plasmas}\ }\textbf {\bibinfo {volume} {21}},\ \bibinfo {pages} {055708}
  (\bibinfo {year} {2014})}\BibitemShut {NoStop}%
\bibitem [{\citenamefont {Evstatiev}\ and\ \citenamefont
  {Shadwick}(2013)}]{Evstatiev13}%
  \BibitemOpen
  \bibfield  {author} {\bibinfo {author} {\bibfnamefont {E.}~\bibnamefont
  {Evstatiev}}\ and\ \bibinfo {author} {\bibfnamefont {B.}~\bibnamefont
  {Shadwick}},\ }\href@noop {} {\bibfield  {journal} {\bibinfo  {journal} {J.
  Comput. Phys.}\ }\textbf {\bibinfo {volume} {245}},\ \bibinfo {pages} {376}
  (\bibinfo {year} {2013})}\BibitemShut {NoStop}%
\bibitem [{\citenamefont {Qin}\ \emph {et~al.}(2016)\citenamefont {Qin},
  \citenamefont {Liu}, \citenamefont {Xiao}, \citenamefont {Zhang},
  \citenamefont {He}, \citenamefont {Wang}, \citenamefont {Sun}, \citenamefont
  {Burby}, \citenamefont {Ellison},\ and\ \citenamefont {Zhou}}]{Qin16SymVM}%
  \BibitemOpen
  \bibfield  {author} {\bibinfo {author} {\bibfnamefont {H.}~\bibnamefont
  {Qin}}, \bibinfo {author} {\bibfnamefont {J.}~\bibnamefont {Liu}}, \bibinfo
  {author} {\bibfnamefont {J.}~\bibnamefont {Xiao}}, \bibinfo {author}
  {\bibfnamefont {R.}~\bibnamefont {Zhang}}, \bibinfo {author} {\bibfnamefont
  {Y.}~\bibnamefont {He}}, \bibinfo {author} {\bibfnamefont {Y.}~\bibnamefont
  {Wang}}, \bibinfo {author} {\bibfnamefont {Y.}~\bibnamefont {Sun}}, \bibinfo
  {author} {\bibfnamefont {J.~W.}\ \bibnamefont {Burby}}, \bibinfo {author}
  {\bibfnamefont {L.}~\bibnamefont {Ellison}}, \ and\ \bibinfo {author}
  {\bibfnamefont {Y.}~\bibnamefont {Zhou}},\ }\href@noop {} {\bibfield
  {journal} {\bibinfo  {journal} {Nuclear Fusion}\ }\textbf {\bibinfo {volume}
  {56}},\ \bibinfo {pages} {014001} (\bibinfo {year} {2016})}\BibitemShut
  {NoStop}%
\bibitem [{\citenamefont {Xiao}\ \emph {et~al.}(2015)\citenamefont {Xiao},
  \citenamefont {Qin}, \citenamefont {Liu}, \citenamefont {He}, \citenamefont
  {Zhang},\ and\ \citenamefont {Sun}}]{Xiao15Pop}%
  \BibitemOpen
  \bibfield  {author} {\bibinfo {author} {\bibfnamefont {J.}~\bibnamefont
  {Xiao}}, \bibinfo {author} {\bibfnamefont {H.}~\bibnamefont {Qin}}, \bibinfo
  {author} {\bibfnamefont {J.}~\bibnamefont {Liu}}, \bibinfo {author}
  {\bibfnamefont {Y.}~\bibnamefont {He}}, \bibinfo {author} {\bibfnamefont
  {R.}~\bibnamefont {Zhang}}, \ and\ \bibinfo {author} {\bibfnamefont
  {Y.}~\bibnamefont {Sun}},\ }\href@noop {} {\bibfield  {journal} {\bibinfo
  {journal} {Physics of Plasmas}\ }\textbf {\bibinfo {volume} {22}},\ \bibinfo
  {pages} {112504} (\bibinfo {year} {2015})}\BibitemShut {NoStop}%
\bibitem [{\citenamefont {He}\ \emph {et~al.}(2015{\natexlab{b}})\citenamefont
  {He}, \citenamefont {Sun}, \citenamefont {Zhou}, \citenamefont {Liu},\ and\
  \citenamefont {Qin}}]{He15KSym}%
  \BibitemOpen
  \bibfield  {author} {\bibinfo {author} {\bibfnamefont {Y.}~\bibnamefont
  {He}}, \bibinfo {author} {\bibfnamefont {Y.}~\bibnamefont {Sun}}, \bibinfo
  {author} {\bibfnamefont {Z.}~\bibnamefont {Zhou}}, \bibinfo {author}
  {\bibfnamefont {J.}~\bibnamefont {Liu}}, \ and\ \bibinfo {author}
  {\bibfnamefont {H.}~\bibnamefont {Qin}},\ }\href@noop {} {\bibfield
  {journal} {\bibinfo  {journal} {arXiv:1509.07794}\ } (\bibinfo {year}
  {2015}{\natexlab{b}})}\BibitemShut {NoStop}%
\bibitem [{\citenamefont {Ruth}(1983)}]{Ruth83}%
  \BibitemOpen
  \bibfield  {author} {\bibinfo {author} {\bibfnamefont {R.~D.}\ \bibnamefont
  {Ruth}},\ }\href@noop {} {\bibfield  {journal} {\bibinfo  {journal} {IEEE
  Trans. Nucl. Sci}\ }\textbf {\bibinfo {volume} {30}},\ \bibinfo {pages}
  {2669} (\bibinfo {year} {1983})}\BibitemShut {NoStop}%
\bibitem [{\citenamefont {Feng}(1985)}]{Feng85}%
  \BibitemOpen
  \bibfield  {author} {\bibinfo {author} {\bibfnamefont {K.}~\bibnamefont
  {Feng}},\ }in\ \href@noop {} {\emph {\bibinfo {booktitle} {the Proceedings of
  1984 Beijing Symposium on Differential Geometry and Differential
  Equations}}},\ \bibinfo {editor} {edited by\ \bibinfo {editor} {\bibfnamefont
  {K.}~\bibnamefont {Feng}}}\ (\bibinfo  {publisher} {Science Press},\ \bibinfo
  {year} {1985})\ pp.\ \bibinfo {pages} {42--58}\BibitemShut {NoStop}%
\bibitem [{\citenamefont {Feng}\ and\ \citenamefont {Qin}(2010)}]{Feng10}%
  \BibitemOpen
  \bibfield  {author} {\bibinfo {author} {\bibfnamefont {K.}~\bibnamefont
  {Feng}}\ and\ \bibinfo {author} {\bibfnamefont {M.}~\bibnamefont {Qin}},\
  }\href@noop {} {\emph {\bibinfo {title} {Symplectic Geometric Algorithms for
  Hamiltonian Systems}}}\ (\bibinfo  {publisher} {Springer-Verlag},\ \bibinfo
  {year} {2010})\BibitemShut {NoStop}%
\bibitem [{\citenamefont {Hairer}\ \emph {et~al.}(2003)\citenamefont {Hairer},
  \citenamefont {Lubich},\ and\ \citenamefont {Wanner}}]{Hairer03}%
  \BibitemOpen
  \bibfield  {author} {\bibinfo {author} {\bibfnamefont {E.}~\bibnamefont
  {Hairer}}, \bibinfo {author} {\bibfnamefont {C.}~\bibnamefont {Lubich}}, \
  and\ \bibinfo {author} {\bibfnamefont {G.}~\bibnamefont {Wanner}},\
  }\href@noop {} {\emph {\bibinfo {title} {Geometric Numerical Integration:
  Structure-Preserving Algorithms for Ordinary Differential Equations}}}\
  (\bibinfo  {publisher} {Springer},\ \bibinfo {address} {New York},\ \bibinfo
  {year} {2003})\BibitemShut {NoStop}%
\bibitem [{\citenamefont {Shang}(1999)}]{Shang99}%
  \BibitemOpen
  \bibfield  {author} {\bibinfo {author} {\bibfnamefont {Z.}~\bibnamefont
  {Shang}},\ }\href@noop {} {\bibfield  {journal} {\bibinfo  {journal} {Numer.
  Math.}\ }\textbf {\bibinfo {volume} {83}},\ \bibinfo {pages} {477496}
  (\bibinfo {year} {1999})}\BibitemShut {NoStop}%
\bibitem [{\citenamefont {Ellison}\ \emph {et~al.}(2015)\citenamefont
  {Ellison}, \citenamefont {Burby},\ and\ \citenamefont
  {Qin}}]{Ellison2015489}%
  \BibitemOpen
  \bibfield  {author} {\bibinfo {author} {\bibfnamefont {C.}~\bibnamefont
  {Ellison}}, \bibinfo {author} {\bibfnamefont {J.}~\bibnamefont {Burby}}, \
  and\ \bibinfo {author} {\bibfnamefont {H.}~\bibnamefont {Qin}},\ }\href
  {\doibase http://dx.doi.org/10.1016/j.jcp.2015.09.007} {\bibfield  {journal}
  {\bibinfo  {journal} {Journal of Computational Physics}\ }\textbf {\bibinfo
  {volume} {301}},\ \bibinfo {pages} {489} (\bibinfo {year}
  {2015})}\BibitemShut {NoStop}%
\bibitem [{\citenamefont {Qiang}\ \emph {et~al.}(2004)\citenamefont {Qiang},
  \citenamefont {Furman},\ and\ \citenamefont {Ryne}}]{Qiang04}%
  \BibitemOpen
  \bibfield  {author} {\bibinfo {author} {\bibfnamefont {J.}~\bibnamefont
  {Qiang}}, \bibinfo {author} {\bibfnamefont {M.}~\bibnamefont {Furman}}, \
  and\ \bibinfo {author} {\bibfnamefont {R.}~\bibnamefont {Ryne}},\ }\href@noop
  {} {\bibfield  {journal} {\bibinfo  {journal} {Journal of Computational
  Physics}\ }\textbf {\bibinfo {volume} {198}},\ \bibinfo {pages} {278}
  (\bibinfo {year} {2004})}\BibitemShut {NoStop}%
\bibitem [{\citenamefont {Feng}\ and\ \citenamefont
  {Shang}(1995{\natexlab{a}})}]{fengs95vpa}%
  \BibitemOpen
  \bibfield  {author} {\bibinfo {author} {\bibfnamefont {K.}~\bibnamefont
  {Feng}}\ and\ \bibinfo {author} {\bibfnamefont {Z.}~\bibnamefont {Shang}},\
  }\href@noop {} {\bibfield  {journal} {\bibinfo  {journal} {Numer. Math.}\
  }\textbf {\bibinfo {volume} {71}},\ \bibinfo {pages} {451} (\bibinfo {year}
  {1995}{\natexlab{a}})}\BibitemShut {NoStop}%
\bibitem [{\citenamefont {Blanes}\ \emph {et~al.}(1999)\citenamefont {Blanes},
  \citenamefont {Casas},\ and\ \citenamefont {Ros}}]{Blanes99}%
  \BibitemOpen
  \bibfield  {author} {\bibinfo {author} {\bibfnamefont {S.}~\bibnamefont
  {Blanes}}, \bibinfo {author} {\bibfnamefont {F.}~\bibnamefont {Casas}}, \
  and\ \bibinfo {author} {\bibfnamefont {J.}~\bibnamefont {Ros}},\ }\href@noop
  {} {\bibfield  {journal} {\bibinfo  {journal} {SIAM J. Sci. Comp.}\ }\textbf
  {\bibinfo {volume} {21}},\ \bibinfo {pages} {711} (\bibinfo {year}
  {1999})}\BibitemShut {NoStop}%
\bibitem [{\citenamefont {Blanes}\ \emph {et~al.}(2006)\citenamefont {Blanes},
  \citenamefont {Casas},\ and\ \citenamefont {Murua}}]{Blanes06}%
  \BibitemOpen
  \bibfield  {author} {\bibinfo {author} {\bibfnamefont {S.}~\bibnamefont
  {Blanes}}, \bibinfo {author} {\bibfnamefont {F.}~\bibnamefont {Casas}}, \
  and\ \bibinfo {author} {\bibfnamefont {A.}~\bibnamefont {Murua}},\
  }\href@noop {} {\bibfield  {journal} {\bibinfo  {journal} {SIAM J. Sci.
  Comp.}\ }\textbf {\bibinfo {volume} {27}},\ \bibinfo {pages} {1817} (\bibinfo
  {year} {2006})}\BibitemShut {NoStop}%
\bibitem [{\citenamefont {Blanes}\ \emph {et~al.}(2010)\citenamefont {Blanes},
  \citenamefont {Diele}, \citenamefont {Marangi},\ and\ \citenamefont
  {Ragni}}]{Blanes10}%
  \BibitemOpen
  \bibfield  {author} {\bibinfo {author} {\bibfnamefont {S.}~\bibnamefont
  {Blanes}}, \bibinfo {author} {\bibfnamefont {F.}~\bibnamefont {Diele}},
  \bibinfo {author} {\bibfnamefont {C.}~\bibnamefont {Marangi}}, \ and\
  \bibinfo {author} {\bibfnamefont {S.}~\bibnamefont {Ragni}},\ }\href@noop {}
  {\bibfield  {journal} {\bibinfo  {journal} {Journal of Computational and
  Applied Mathematics}\ }\textbf {\bibinfo {volume} {235}},\ \bibinfo {pages}
  {646} (\bibinfo {year} {2010})}\BibitemShut {NoStop}%
\bibitem [{\citenamefont {Feng}\ and\ \citenamefont
  {Shang}(1995{\natexlab{b}})}]{Feng95vpa}%
  \BibitemOpen
  \bibfield  {author} {\bibinfo {author} {\bibfnamefont {K.}~\bibnamefont
  {Feng}}\ and\ \bibinfo {author} {\bibfnamefont {Z.}~\bibnamefont {Shang}},\
  }\href@noop {} {\bibfield  {journal} {\bibinfo  {journal} {Numer. Math.}\
  }\textbf {\bibinfo {volume} {71}},\ \bibinfo {pages} {451} (\bibinfo {year}
  {1995}{\natexlab{b}})}\BibitemShut {NoStop}%
\bibitem [{\citenamefont {McLachlan}\ and\ \citenamefont
  {Quispel}(2002)}]{Mclachlan02}%
  \BibitemOpen
  \bibfield  {author} {\bibinfo {author} {\bibfnamefont {R.~I.}\ \bibnamefont
  {McLachlan}}\ and\ \bibinfo {author} {\bibfnamefont {G.~R.~W.}\ \bibnamefont
  {Quispel}},\ }\href@noop {} {\bibfield  {journal} {\bibinfo  {journal} {Acta
  Numer.}\ }\textbf {\bibinfo {volume} {11}},\ \bibinfo {pages} {341} (\bibinfo
  {year} {2002})}\BibitemShut {NoStop}%
\bibitem [{\citenamefont {Yoshida}(1990)}]{Yoshida90}%
  \BibitemOpen
  \bibfield  {author} {\bibinfo {author} {\bibfnamefont {H.}~\bibnamefont
  {Yoshida}},\ }\href@noop {} {\bibfield  {journal} {\bibinfo  {journal} {Phys.
  Lett. A.}\ }\textbf {\bibinfo {volume} {150}},\ \bibinfo {pages} {262}
  (\bibinfo {year} {1990})}\BibitemShut {NoStop}%
\bibitem [{\citenamefont {Suzuki}(1992)}]{Suzuki92}%
  \BibitemOpen
  \bibfield  {author} {\bibinfo {author} {\bibfnamefont {M.}~\bibnamefont
  {Suzuki}},\ }\href@noop {} {\bibfield  {journal} {\bibinfo  {journal} {Phys.
  Lett. A}\ }\textbf {\bibinfo {volume} {165}},\ \bibinfo {pages} {387}
  (\bibinfo {year} {1992})}\BibitemShut {NoStop}%
\bibitem [{\citenamefont {McMillan}(1950)}]{McMillan50}%
  \BibitemOpen
  \bibfield  {author} {\bibinfo {author} {\bibfnamefont {E.~M.}\ \bibnamefont
  {McMillan}},\ }\href@noop {} {\bibfield  {journal} {\bibinfo  {journal}
  {Phys. Rev.}\ }\textbf {\bibinfo {volume} {79}},\ \bibinfo {pages} {498}
  (\bibinfo {year} {1950})}\BibitemShut {NoStop}%
\end{thebibliography}%

\end{document}